\documentclass[10pt,journal]{IEEEtran}
\usepackage{amssymb}
\usepackage{mathrsfs}
\usepackage{amssymb}
\usepackage{amsmath}
\usepackage{mathtools}
\usepackage{bm}
\usepackage{epsfig}
\usepackage{color}
\usepackage{csquotes}
\usepackage{caption}
\usepackage{subcaption}
\usepackage{enumitem}
\usepackage{gensymb}
\usepackage{float}
\usepackage[normalem]{ulem}
\usepackage{xcolor}
\usepackage{accents}
\usepackage{cite}
\newcommand\redst{\bgroup\markoverwith{\textcolor{red}{\rule[0.5ex]{2pt}{1.0pt}}}\ULon}

\newcommand\PsD{\mathbf{P}_{\hspace{-2pt}\scriptscriptstyle{D}}}

\newtheorem{assumption}{Assumption}
\newtheorem{theorem}{Theorem}
\newtheorem{lemma}{Lemma}
\newtheorem{proposition}{Proposition}
\newtheorem{definition}{Definition}
\newtheorem{remark}{Remark}
\DeclareMathOperator*{\argmin}{arg\,min}
\DeclareMathOperator*{\argmax}{arg\,max}

\usepackage[ruled,vlined]{algorithm2e}
\SetArgSty{textnormal}
\SetKwInput{KwResult}{Initialization}
\usepackage[labelformat=simple]{subcaption}

\begin{document}

\title{Jacobi-Style Iteration for Distributed \\Submodular Maximization}
\author{Bin~Du$^{\dagger}$,
        Kun~Qian$^{\ddagger}$,
        Christian~Claudel$^{\ddagger}$,
        and~Dengfeng Sun$^{\dagger}$
\thanks{$^{\dagger}$Bin Du is Ph.D. student, and Dengfeng Sun is Associate Professor, with the School of Aeronautics and Astronautics, Purdue University, West~Lafayette,~IN 47907, {\tt\small \{du185,dsun\}@purdue.edu}}%
\thanks{$^{\ddagger}$Kun Qian is Ph.D. student, and Christian Claudel is Assistant Professor, with the Department  of Civil, Architectural, and Environmental Engineering, University of Texas at Austin, Austin, TX 78712, {\tt\small \{kunqian,christian.claudel\}@utexas.edu }}%
\thanks{Bin Du and Kun Qian contributed equally to this manuscript.}
}
\maketitle

\begin{abstract}
This paper presents a novel Jacobi-style iteration algorithm for solving the problem of distributed submodular maximization, in which each agent determines its own strategy from a finite set so that the global submodular objective function is jointly maximized. Building on the multi-linear extension of the global submodular function, we expect to achieve the solution from a probabilistic, rather than deterministic, perspective, and thus transfer the considered problem from a discrete domain into a continuous domain. Since it is observed that an unbiased estimation of the gradient of multi-linear extension function~can be obtained by sampling the agents' local decisions, a projected stochastic gradient algorithm is proposed to solve the problem. Our algorithm enables the distributed updates among all individual agents and is proved to asymptotically converge to a desirable equilibrium solution. Such an equilibrium solution is guaranteed to achieve at least $1/2$-suboptimal bound, which is comparable to the state-of-art in the literature. Moreover, we further enhance the proposed algorithm by handling the scenario in which agents' communication delays are present. The enhanced algorithmic framework admits a more realistic distributed implementation of our approach. Finally, a movie recommendation task is conducted on a real-world movie rating data set, to validate the numerical performance of the proposed algorithms.
\end{abstract}

\section{Introduction}\label{sec:intro}

In this paper, we focus on the distributed maximization of submodular functions, involving a network of $I$ agents, which aims at cooperatively solving the following problem,
\begin{equation}\label{submodularmax}
\begin{aligned}
\mathop {\text{maximize} } \quad &F(a_1, a_2, \cdots, a_I)\\
{\text{subject to} } \quad & a_i \in \mathcal{A}_i,\; i = 1,2,\cdots,I.
\end{aligned} \tag{P}
\end{equation}
In problem~(\ref{submodularmax}), each agent $i \in \mathcal{I} := \{1,2,\cdots,I\}$ within the network is expected to select a desirable local strategy $a_i$ from the private finite set~$\mathcal{A}_i$, such that the common objective function $F: \mathcal{A}_1 \times\mathcal{A}_2 \times\cdots \times \mathcal{A}_I \to \mathbb{R}$ is jointly maximized. Here, we assume that the objective function is submodular and additionally monotone; see definitions in Section~\ref{sec:prob}. In fact, such a~(distributed) monotone submodular maximization problem has gained increasing attention in recent years, primarily due to the fact that it can be widely adopted in numerous applications, including resource allocation~\mbox{\cite{marden2016role,streeter2009online}}, sensor placement \cite{krause2008near,jawaid2015submodularity}, data summarization \cite{mirzasoleiman2013distributed,badanidiyuru2014streaming}, information gathering \cite{atanasov2015decentralized,schlotfeldt2018anytime}, to name a few.

{While the submodular maximization problem has successfully found so many applications, solving problem~(\ref{submodularmax}) is known to be NP-hard \cite{lovasz1983submodular}, even from a standard centralized perspective. Due to this fact, the approximation methods which are able to guarantee suboptimal solutions are broadly studied in the literature~\cite{sviridenko2004note,buchbinder2015tight,nemhauser1978analysis,fisher1978analysis,calinescu2011maximizing}. Among these various approximation methods, the greedy algorithm \cite{nemhauser1978analysis,fisher1978analysis,calinescu2011maximizing} attracts the most attentions by researchers. The key idea of this greedy algorithm is to determine the single best strategy at each time, by maximizing the marginal gain of the submodular function. It is shown in \cite{fisher1978analysis} that an $1/2$-suboptimal solution can be guaranteed by the greedy algorithm, i.e., the obtained solution $A^{g}$ is ensured to have $F(A^g) \ge 1/2\cdot F(A^\star)$ where $A^\star$ represents the global optimal solution. In particular, when the considered submodular maximization problem has the constraints of some specific forms \cite{nemhauser1978analysis,calinescu2011maximizing}, the suboptimal bound can be further improved to $1-1/e$ with $e$ being the natural constant. Nevertheless, it should be remarked that the greedy algorithm is inherently a sequential updating scheme, since the single best strategy needs to be determined one by one. On this account, the greedy algorithm may not be useful or even infeasible in many applications, especially when a large number of agents are involved in the problem.

In order to address such a sequential updating issue, recent papers \cite{gharesifard2017distributed,grimsman2018impact,sun2020distributed} have developed distributed variants of the greedy algorithm in which the best single strategies can be determined simultaneously in a parallel architecture. However, as shown in \cite{grimsman2018impact}, when it comes to a distributed setting (with limited information), the suboptimal bound needs to be degraded from $1/2$ to $1/(1+\beta)$ where $\beta$ is a constant related to the multi-agent network topology. Although the best network topology is further designed in \cite{sun2020distributed} to enhance performance of the distributed algorithm, the fact that $\beta \ge 1$ makes the obtained solution worse than $1/2$-suboptimal in nature.

It is worthy to note that some other distributed approaches have also been devised to solve the problem of maximizing submodular functions. The authors in \cite{qu2019distributed} develop a new distributed method while considering the application of multi-agent task assignment. It is proved that the obtained solution is at least $1/2$-suboptimal. Moreover, the distributed submodular maximization problem is studied in both discrete and continuous settings \cite{mokhtari2018decentralized}, and the proposed algorithms are guaranteed to converge asymptotically to the $1-1/e$ suboptimal bound. A similar algorithm is also developed in \cite{xie2019decentralized} which further improves convergence performance. We remark that the problem considered in our paper is significantly different from the ones in \cite{qu2019distributed,mokhtari2018decentralized,xie2019decentralized}, where a separable structure of the global objective function is specifically assumed. Precisely, in their problem setups \cite{qu2019distributed,mokhtari2018decentralized,xie2019decentralized}, each agent (or task) maintains a local utility function, and the global objective is to optimize the summation of local functions. This inherently makes the problem easier to solve. For instance, given that the global gradient is computed by summing all local gradients due to the specific structure of the function, the desired global information can be achieved by a standard consensus procedure as in~\cite{mokhtari2018decentralized}, or more directly, the technique of gradient tracking as in~\cite{xie2019decentralized}. However, in our problem, such an idea is not applicable due to the generality of the global objective function.


To sum up, while the $1/2$-suboptimal bound is known to be the best result that one can achieve in general, there is no existing distributed algorithm yet, which guarantees such a bound when solving problem~(\ref{submodularmax}).} Motivated by this, it is exactly the purpose of this paper to devise such a distributed algorithm. Our contributions are summarized as follows. A novel Jacobi-style algorithm is proposed for solving the problem of distributed submodular maximization. Unlike existing works which are based on the greedy algorithm, we start from a probabilistic perspective, build on the multi-linear extension of the submodular function, and eventually transfer the considered problem from a discrete domain into a continuous domain. 
By leveraging the fact that an unbiased estimation of the gradient can be achieved by simply sampling the agents' local decisions, we develop a projected stochastic gradient algorithm. It is proved that our algorithm converges to an equilibrium solution which is guaranteed to be at least $1/2$-suboptimal. In addition, by handling the scenario where communication delays are present among agents, we further enhance the proposed algorithm to be implementable in a more realistic distributed architecture. The same convergence performance is proved for the enhanced distributed algorithm. Finally, the movie recommendation task is conducted on a real-world movie rating data set, and the simulation results validates the effectiveness of our algorithms.

The remainder of this paper is organized as follows. Sec.~\ref{sec:prob} formally defines the considered distributed submodular maximization problem. Sec.~\ref{sec:SGD} develops the projected stochastic gradient algorithm, and Sec. \ref{sec:DecentralizedAlgo} further enhances the proposed algorithm by dealing with the communication delays. Numerical simulations are presented in Sec.~\ref{sec:simulation}. Lastly, Sec.~\ref{sec:conclusion} concludes this paper. For the reader’s convenience, the proofs of propositions and theorems are provided in Appendix.

\section{Problem Statement}\label{sec:prob}

Let us first formalize the considered distributed submodular maximization problem. For the sake of notational simplicity, we here assume that each agent's finite set $\mathcal{A}_i$ has the same size $K$, i.e., $|\mathcal{A}_i| = K, \,\forall i \in \mathcal{I}$. In addition, we stack all agents' local strategies as a vector $A = [a_1,a_2, \cdots,a_I]^\top \in \mathcal{A}$, where the entire searching space $\mathcal{A}$ is the Cartesian product of $\mathcal{A}_i$'s, i.e., $ \mathcal{A}:= \prod_{i\in\mathcal{I}}\mathcal{A}_i$ and $|\mathcal{A}| = K^I$. Based on the set of collected strategies, the objective function in problem (\ref{submodularmax}) can be succinctly written as $F(A)$. Note that we allow each agent to choose the empty set as its strategy, i.e., $a_i = \emptyset$; and in particular, let $F(\emptyset) = 0$. With slight abuse of notations, we say that the set of strategies $A'$ is contained in~$A$, denoted as $A' \subseteq A$, if some component $a_i'$ in $A'$ has $a_i' = \emptyset$ and other components have $a_j' = a_j,\,\forall j \neq i$. In this case, we also denote $A = A' \cup \{a_i\}$. Furthermore, we restrict the objective function $F(A)$ to satisfy the following assumption.
\begin{assumption}\label{assumption:function}
  The function $F(A)$ is assumed to satisfy:
  \begin{enumerate}[label=(A.{{\arabic*}}),leftmargin=3\parindent]
     \item (\textit{Monotone}) If two sets $A'$ and $A$ have $A' \subseteq A$, then it implies $F(A') \le F(A)$.
     \item (\textit{Submodular}) If two sets $A'$ and $A$ have $A' \subseteq A$, then $F(A' \cup\{a\}) - F(A') \ge F(A\cup\{a\}) - F(A)$ for any~$a$.
   \end{enumerate} 
\end{assumption}

Clearly, the ultimate goal of problem~(\ref{submodularmax}) is to find the group of optimal strategies $A^\star = [a_1^\star,a_2^\star, \cdots,a_I^\star]^\top \in \mathcal{A}$ such that $F(A^\star)$ gives the maximal function value against all other possible groups of strategies. However, as mentioned before in~Section~\ref{sec:intro}, achieving such a goal is NP-hard in general. The challenges mainly come from the following two aspects. First, the finite set $\mathcal{A}_i$ from which the agent chooses its strategy is inherently discrete. Thus, the well-developed techniques of continuous optimization cannot be adopted for solving the problem. Although the bright side of this fact is that one can apply some search-tree based approaches since the set $\mathcal{A}_i$ is anyway finite, the computational demand during such a searching procedure is often costly or even infeasible, especially when each individual searching space $\mathcal{A}_i$ is large-scale. This is also related to the second aspect of the challenges. Note that the objective function $F(A)$ can be evaluated only when all individual agents have decided their own strategies. That is to say, the function $F(A)$ mixes the decisions of each agent within the network. In this sense, the size of the entire searching space $\mathcal{A}$ also grows exponentially with respect to the number of agents. This undoubtedly prohibits the idea of using searching procedures to find the joint optimal strategies $A^\star$, when a large number of agents are involved in the problem.

In order to address the above two challenges, in this paper, our ideas are: 1) utilizing the multi-linear extension of the function $F(A)$; see Section~\ref{subsec:multilinear}, and transferring the considered problem into a continuous domain, so that the techniques of continuous optimization can be exploited; and 2) developing the Jacobi-style iteration to decompose the mixing of individual agents' decisions. In particular, the algorithmic framework of our Jacobi-style iteration for each agent $i$ can be abstracted as the following mapping $\mathcal{M}_i: \mathcal{A}_1 \times\mathcal{A}_{2} \times\cdots \times \mathcal{A}_I \to \mathcal{A}_i$ such that
\begin{align}\label{JacobiFramework}
  a_i^+ = \mathcal{M}_i\Big( a_i^-,A^r_{-i}\Big).
\end{align}
Here, $a_i^-$ is the $i$-th agent's previous decision of the desired strategy; $A^r_{-i} = [a^r_1, \cdots, a^r_{i-1}, a^r_{i+1} \cdots a^r_I]^\top \in \mathcal{A}_{-i}$ is the collection of the received decisions which have been made by the other agents~$j \neq i$; and $a_i^+$ is the $i$-th agent's updated decision based on the previous $a_i^-$ and the received information $a_j^r$'s. It should be emphasized that, under such a framework (\ref{JacobiFramework}), each agent only needs to take charge in its own decision, by receiving the information $a_j^r$ from other agents. Thus, the computational complexity is expected to be primarily reduced, compared to the aforementioned searching procedure. In addition, another advantage of the framework is that individual agents can perform the update of decisions simultaneously, so that the overall processing time can be further saved. However, it is also worthy to note that there are two potential issues regarding the framework. First, the mapping~$\mathcal{M}_i$ suggests that an {instantaneous} all-to-all communication is required, i.e., each agent~$i$ needs to communicate with all other agents to receive the most updated information~$a_j^r$'s. Second, it is not clear that what kind of solution will be produced by the iteration~(\ref{JacobiFramework}). {For the first concern, we here remark that the communication requirement will be eliminated later on; see Section~\ref{sec:DecentralizedAlgo}, so that our algorithm is implementable in a distributed architecture with communication delays.} For the second one, we are interested in finding an equilibrium solution~$A^e$ which is formally defined as below. Interestingly, it can be proved that such an equilibrium is guaranteed to be at least $1/2$-suboptimal which is comparable to the state-of-art in the literature; see Section~\ref{sec:intro}. Before defining the equilibrium~$A^e$, let us first introduce another assumption related to the objective function $F(A)$.

\begin{assumption}[Maximum Distinguishable]\label{assump:distinguishable}
It is assumed that, once other agents $j \neq i$ have decided their strategies $A_{-i}$, the $i$-th agent's best strategy $a_i$, which gives the maximum function value $F(a_i; A_{-i})$, is unique.
\end{assumption}

Here, to concentrate on the effect of strategy $a_i$ on the function value, $F(A)$ is expressed as the specific form $ F(a_i; A_{- i})$. {We note that the above Assumption~\ref{assump:distinguishable} is easily satisfied in many applications, especially when some certain randomness is involved in the function values.} In particular, we denote $\Delta^\text{max}$ the maximum discrepancy of function values between two individual strategies $a_i$ and $a_i'$ when other strategies $A_{-i}$ have been fixed, i.e.,
\begin{align}\label{deltaMax}
   \Delta^\text{max} = \max_{A_{-i}\in \mathcal{A}_{-i}}\big\{ \max_{a_i, a'_i \in \mathcal{A}_i} \{F(a_i; A_{-i})- F(a'_i;A_{-i}\}\big\}.
\end{align} 
It is trivial to see that $\Delta^\text{max}$ has to be strictly greater than zero due to Assumption~\ref{assump:distinguishable}. Next, we formalize the definition of the equilibrium $A^e$.

\begin{definition}\label{def:equilibrium}
  A solution $A^e = [a_1^e,a_2^e, \cdots,a_I^e]^\top \in \mathcal{A}$ is said to be the equilibrium to problem~(\ref{submodularmax}), if and only if it satisfies the following condition:
  \begin{align}\label{eq:equilibrium}
    a_i^e = \argmax_{a_i \in \mathcal{A}_i} \; F(a^e_1, \cdots\hspace{-2pt}, a^e_{i-1}, a_i,a^e_{i+1}, \cdots\hspace{-2pt}, a^e_I),\; i \in \mathcal{I}.
  \end{align}
\end{definition}
Remark that the uniqueness of $a_i^e$ in (\ref{eq:equilibrium}) is guaranteed by the maximum distinguishable assumption of the objective function~$F(A)$. Moreover, by the definition of the equilibrium solution, it is clear that $A^e$ is not unique for problem~(\ref{submodularmax}) and $A^\star$ is just a specific equilibrium which has the maximal function value. Next, we show, by the following proposition, that any equilibrium $A^e$ satisfying (\ref{eq:equilibrium}) must be at least $1/2$-suboptimal to our problem.

\begin{proposition}\label{prop:suboptimal}
  Suppose that the function $F(A)$ satisfies the conditions in Assumption~\ref{assumption:function}. Let $A^\star$ be an optimal solution to problem (\ref{submodularmax}) and $A^e$ be an equilibrium solution following Definition~\ref{def:equilibrium}, then it holds that
  \begin{align}
    F(A^e) \ge \frac{1}{2}\cdot F(A^\star).
  \end{align}
\end{proposition}
\begin{IEEEproof}
  See Appendix A.
\end{IEEEproof}

\section{Stochastic Gradient Based Method}\label{sec:SGD}

To solve for the equilibrium solution $A^e$, we develop a stochastic gradient based solution method in this section. As an important building block of our method, the multi-linear extension of the function $F(A)$ is first introduced.

\subsection{Multi-Linear Extension}\label{subsec:multilinear}

Let us recall that $A = [a_1,a_2, \cdots,a_I]^\top$ is the set of $I$ local strategies in which each $a_i$ is chosen from the finite set~$\mathcal{A}_i$. Now, instead of expecting the individual agents to seek the desired deterministic strategies from $\mathcal{A}_i$'s, we assign each agent a discrete probability distribution $\mathbf{p}_i = [p_i(a_i)]_{a_i \in \mathcal{A}_i} \in [0,1]^K$, where $p_i(a_i)$ represents the probability of choosing $a_i$ as the \mbox{$i$-th} agent's strategy. On this account, we define the multi-linear extension of $F(A)$ as the function $f({P}): [0,1]^{I\cdot K} \to \mathbb{R}$,
\begin{align}\label{multilinear}
  f(P) = \sum_{a_1 \in \mathcal{A}_1} p_1(a_1)\sum_{a_2 \in \mathcal{A}_2} p_2(a_2) \cdots\sum_{a_I \in \mathcal{A}_I} p_I(a_I) \cdot F ({A}),
\end{align}
where the argument $P$ is a compact vector which stacks all $\mathbf{p}_i$'s, i.e., $P = [\mathbf{p}_1^\top, \mathbf{p}_2^\top, \cdots, \mathbf{p}_I^\top]^\top \in [0,1]^{I\cdot K}$. We shall emphasize that a core property of the multi-linear extension~(\ref{multilinear}) is its natural connection to the original function $F(A)$. That~is,
\begin{align}\label{exp}
  f(P) = \mathbb{E}_{\tilde{\mathbf{a}}_i \sim \mathbf{p}_i, i \in \mathcal{I}} \big[F(\tilde{\mathbf{A}})\big],
\end{align}
where the expectation is taken from $\tilde{\mathbf{A}} = [\tilde{\mathbf{a}}_1,\tilde{\mathbf{a}}_2, \cdots,\tilde{\mathbf{a}}_I]^\top$ and each $\tilde{\mathbf{a}}_i$ is an independent random variable following the discrete distribution $\mathbf{p}_i$.
Moreover, consider that each $\mathbf{p}_i$ has to be subject to $p_i(a_i) \ge 0$ and $\mathbf{1}^\top\mathbf{p}_i = 1$, let us express those constraints as the following probability simplex $\mathcal{S}$, i.e.,
\begin{align}
\mathbf{p}_i \in \mathcal{S}: =\Big\{\mathbf{p} \,|\,\mathbf{1}^\top \mathbf{p} = 1,\, \mathbf{p} \in [0,1]^K\Big\}.
\end{align}
In particular, we say $\mathbf{p}_i$ is a vertex of the simplex $\mathcal{S}$ if it has $\|\mathbf{p}_i\|_{\infty} = 1$. i.e., there is exactly one component $p_i(a_i)$ which equals one and all others are zeros.

With the help of this multi-linear extension function $f(P)$, our goal now becomes to seek the desired probability distribution $\mathbf{p}_i$'s such that $f(P)$ is optimized. Therefore, the submodular maximization problem can be equivalently written~as,
\begin{equation}\label{prob_max}
\begin{aligned}
\mathop {\text{maximize} } \quad &f(P) = \mathbb{E}_{\tilde{\mathbf{a}}_i \sim \mathbf{p}_i, i \in \mathcal{I}} \big[F(\tilde{\mathbf{A}})\big]\\
{\text{subject to} } \quad & \mathbf{p}_i \in \mathcal{S}, \; i \in \mathcal{I}.
\end{aligned}
\end{equation}
Recall that, in the original problem~(\ref{submodularmax}), we are interested in seeking the equilibrium solution $A^e$ which is defined by Definition~\ref{def:equilibrium}. Now, following the same path, we introduce a similar equilibrium solution $P^e$ to problem (\ref{prob_max}), based on the defined $A^e$.
\begin{definition}\label{def:Pe}
  A solution $P^e = [{\mathbf{p}^e_1}^\top, {\mathbf{p}^e_2}^\top, \cdots\hspace{-2pt}, {\mathbf{p}^e_I}^\top]^\top$ where each $\mathbf{p}_i^e$ is a vertex of the probability simplex, i.e., there exists $a_i^e$ such that $p_i^e(a_i^e) = 1$ and $p_i^e(a_i) = 0$ for $\forall a_i \neq a_i^e$, is said to be an equilibrium, if and only if $A^e = [a_1^e,a_2^e, \cdots,a_I^e]^\top$ is an equilibrium to problem~(\ref{submodularmax}).
\end{definition}

In fact, combining the above Definition \ref{def:equilibrium} and \ref{def:Pe} together establishes the equivalence between problem~(\ref{submodularmax}) and (\ref{prob_max}). In other words, the desired equilibrium $A^e$ to problem~(\ref{submodularmax}) can be easily resulted from the solution $P^e$ by solving problem~(\ref{prob_max}). Next, we develop the projected stochastic gradient algorithm to solve for the equilibrium solution $P^e$.

\subsection{Projected Stochastic Gradient Algorithm}\label{subsec:algorithm}

 Before proceeding to the development of our algorithm, let us first investigate the gradient of function $\nabla f(P)$. Recall that the  function $f(P)$ is a multi-linear extension of $F(A)$ and can be expressed as (\ref{multilinear}), thus the gradient of $f(P)$ with respect to each single component $p_i(a_i)$ can be represented as
\begin{equation}\label{fullGrad}
  \begin{aligned}
      &\nabla_{p_i(a_i)} f (P) = \sum_{a_1 \in \mathcal{A}_1} p_1(a_1)\cdots\sum_{a_{i-1} \in \mathcal{A}_{i-1}} p_{i-1}(a_{i-1})\\
  &\hspace{40pt}\sum_{a_{i+1} \in \mathcal{A}_{i+1}}\hspace{-6pt} p_{i+1}(a_{i+1}) \cdots\sum_{a_I \in \mathcal{A}_I} p_I(a_I)\cdot F (a_i;{A}_{- i}).
  \end{aligned}
\end{equation}
Since $F(A)$ has its expectation interpretation as shown in~(\ref{exp}), the gradient $\nabla_{p_i(a_i)} f (P)$ can be also expressed as the following expectation form,
\begin{align}\label{expGrad}
  \nabla_{p_i(a_i)} f (P) &=\mathbb{E}_{\tilde{\mathbf{a}}_j \sim \mathbf{p}_j, j\neq i}\big[F (a_i;\tilde{{\mathbf{A}}}_{- i})\big].
\end{align}

We remark that, to evaluate the gradient $\nabla_{p_i(a_i)} f (P)$ for the $i$-th agent in the form of (\ref{fullGrad}), it is required to sum all possibilities that are governed by the probability distribution $\mathbf{p}_j$'s for all other agents $j \neq i$. However, due to its expectation form~(\ref{expGrad}), a key observation here is that an unbiased estimation of the gradient can be obtained by sampling the strategies  $a_j$ based on $\mathbf{p}_j$ for $\forall j \neq i$. In this sense, we call $\nabla_{p_i(a_i)} f (P)$ the full gradient which is computed by~(\ref{fullGrad}), and meanwhile denote the following $\nabla_{p_i(a_i)} \hat{f} (P)$ as the sampled stochastic gradient with the sample-size $M \in \mathbb{N}_+$,
\begin{align}\label{stoGrad}
   &\nabla_{p_i(a_i)} \hat{f} (P) = \frac{1}{M} \cdot \sum_{s=1}^M F(a_i; \hat{{A}}_{- i}^s),
\end{align}
where $\hat{A}_{- i}^s = [\hat{a}_1^s, \cdots\hspace{-2pt}, \hat{a}_{i-1}^s, \hat{a}_{i+1}^s, \cdots\hspace{-2pt}, \hat{a}_I^s]^\top$ and each $\hat{a}_j^s, j\neq i$ is the independent and identically distributed (\textit{i.i.d.}) sampled strategy based on the probability distribution $\mathbf{p}_j$.
Taking advantage of this stochastic gradient $\nabla_{p_i(a_i)} \hat{f} (P)$, our projected stochastic gradient algorithm performs the following iteration, with index $k \in \mathbb{N}_+$,
\begin{align}\label{SGD}
  \mathbf{p}_i^{k+1} = \Pi_{\mathcal{S}}\big( \mathbf{p}_i^{k} + \gamma\cdot\nabla_{\mathbf{p}_i} \hat{f}({{P}}^k) \big),\; \forall i \in \mathcal{I}.
\end{align}
In (\ref{SGD}), $P^k$ is the collection of $\mathbf{p}_i^k$'s for all agents at the $k$-th iteration, i.e., $P^k = [{\mathbf{p}^k_1}^\top, {\mathbf{p}^k_2}^\top, \cdots, {\mathbf{p}^k_I}^\top]^\top$; $\nabla_{\mathbf{p}_i} \hat{f}({{P}}^k)$ is a vector which stacks the stochastic gradients for all \mbox{$a_i \in \mathcal{A}_i$}, i.e., $\nabla_{\mathbf{p}_i} \hat{f}({{P}}^k) = [\nabla_{p_i(a_i)} \hat{f} (P^k)]_{a_i \in \mathcal{A}_i}$; $\gamma > 0$ is a constant step-size; and the operator $\Pi_\mathcal{S}(\cdot) : \mathbb{R}^K \to \mathcal{S}$ defines the projection on the probability simplex $\mathcal{S}$, i.e.,
  \begin{align}\label{projection}
    \Pi_{\mathcal{S}}(\mathbf{p}) := \argmin_{\mathbf{x} \in \mathcal{S}}\; \|\mathbf{x} - \mathbf{p}\|^2.
  \end{align}

Our ultimate goal here is to drive the sequence of $P^k$ generated by the iteration~(\ref{SGD}) to the desired equilibrium solution $P^e$ as defined in Definition~\ref{def:Pe}. Before proceeding to the convergence analysis of our algorithm, let us first show, by the following proposition, an alternative way to characterize the equilibrium solution $P^e$.

\begin{proposition}\label{prop:gradMapping}
 Under Assumption~\ref{assump:distinguishable}, the probability distribution $P^e$ is an equilibrium solution to problem~(\ref{prob_max}); see Definition~\ref{def:Pe}, if and only if the following condition is satisfied,
  \begin{align}\label{gradMapping}
    \mathbb{E} \Big[\big\|\mathbf{p}_i^e - \Pi_{\mathcal{S}}\big( \mathbf{p}_i^{e} + \gamma \cdot \nabla_{\mathbf{p}_i} \hat{f}({{P}}^e) \big)\big\|^2\Big] = 0,\; \forall i \in \mathcal{I}.
  \end{align}
\end{proposition}

With the help of the above proposition, we are now in the position to analyze the convergence of our projected stochastic gradient algorithm.
\begin{theorem}\label{theorem:SGD}
  Suppose that Assumption~\ref{assump:distinguishable} is satisfied, and let $\{P^k\}_{k \in \mathbb{N}_+}$ be the sequence generated by the iteration~(\ref{SGD}) with a small enough constant step-size $\gamma$ and a large enough constant sample-size $M$. Then, it holds that
  \begin{align}
  \lim_{k \to \infty} \mathbb{E} \Big[\big\|\mathbf{p}_i^k - \Pi_{\mathcal{S}}\big( \mathbf{p}_i^{k} + \gamma \cdot \nabla_{\mathbf{p}_i} \hat{f}({{P}}^k) \big)\big\|^2\Big] = 0 ,\; \forall i \in \mathcal{I},
  \end{align}
  and furthermore, the running average converges at the rate of $\mathcal{O}(1/T)$ where $T$ is the number of iterations, i.e., there exists a constant $\eta_1 > 0$ such that
  \begin{align}
    \frac{1}{T}&\sum_{k=0}^T \mathbb{E} \Big[\big\|\mathbf{p}_i^k - \Pi_{\mathcal{S}}\big( \mathbf{p}_i^{k} + \gamma \cdot \nabla_{\mathbf{p}_i} \hat{f}({{P}}^k) \big)\big\|^2\Big] \le \frac{\eta_1}{T}.
  \end{align}
\end{theorem}

Note that the proofs of both Proposition~\ref{prop:gradMapping} and Theorem~\ref{theorem:SGD} are provided in Appendix B and C, respectively; in addition, the detailed conditions of the step-size $\gamma$ and sample-size $M$ are also specified in the proof. Now, combining the above Proposition~\ref{prop:gradMapping} and Theorem~\ref{theorem:SGD} together, it has been shown that the projected stochastic gradient algorithm converges to the desired equilibrium $P^e$. To sum up, we outline our scheme as the following Algorithm~\ref{algo:SGD} and provide a few remarks on it.

\begin{algorithm}
 \SetAlgoLined
 \caption{Projected Stochastic Gradient Algorithm}
\vspace{5pt}
  \KwResult{\parbox{\dimexpr\textwidth+2\algomargin\relax}{Each agent $i$ initializes its own probability} \\
  {\parbox{\dimexpr\textwidth-2\algomargin\relax}{distribution $\mathbf{p}_i^0$ (not vertex), samples a set of $M$ strategies}}\\
  \parbox{\dimexpr\textwidth-2\algomargin\relax}{$[\hat{a}_i^s]_{1\le s\le M}$ based on $\mathbf{p}_i^0$, and sends it to all other agents.}\\
  \parbox{\dimexpr\textwidth-2\algomargin\relax}{Set the maximum iteration $K$, and initialize index $k=0$.}
  } 
  \vspace{5pt}
  \While{$0 \le k \le K$ is satisfied}{
  \vspace{5pt}
  Each sensor $i$ \textbf{simultaneously} does\\
   \texttt{(S.1)} Receive the sampled strategies $\hat{a}_j^s$ from all\\ 
   \hspace{34pt}other agents $j$, and evaluate the stochastic \\
   \hspace{34pt}gradient as (\ref{stoGrad});

   \texttt{(S.2)} Update the distribution $\mathbf{p}_i^{k+1}$ as (\ref{SGD});

   \texttt{(S.3)} Sample the $M$ strategies $\hat{a}_i^s$'s based on the \\
   \hspace{34pt}updated $\mathbf{p}_i^{k+1}$, and send it to other agents;

   \texttt{(S.4)} Let $k \leftarrow k+1$, and continue.
   }
   \label{algo:SGD}
\end{algorithm}

\begin{remark}
  It is worth noting that the multi-linear extension function $f(P)$ is neither convex nor concave, thus the considered problem (\ref{prob_max}) belongs to the category of nonconvex optimization. Besides, our problem is also inherently nonsmooth, since the probability simplex constraints are present. In order to achieve the exact convergence for solving such nonsmooth nonconvex optimization  (normally to the stationary point), typical stochastic gradient based approaches follow two paths: 1) increasing the sample-size with iteration numbers~\cite{ghadimi2016mini}; and 2) applying the techniques of variance reduction \cite{reddi2016proximal,li2018simple}. One major novelty of our algorithm herein is the provable exact convergence with a constant \mbox{sample-size}. This primarily benefits from the fact that the stochastic gradient is not only bounded but also has finite possibilities in our problem.
\end{remark}

\begin{remark}
  We also remark that our algorithm is closely related to the well-known EXP3 algorithm\mbox{\cite{auer1995gambling, stoltz2005incomplete, bubeck2012towards}} for solving the multi-armed bandit problems. In fact, the iteration~(\ref{SGD}) of our algorithm can be equivalently rewritten as
  \begin{equation}\label{argmin_SGD}
      \mathbf{p}^{k+1}_i = \argmin_{\mathbf{x} \in \mathcal{S}} \big\{ -\langle \nabla_{\mathbf{p}_i} \hat{f}({{P}}^k),\, \mathbf{x} \rangle + \frac{1}{2\gamma}\| \mathbf{x} - \mathbf{p}^{k}_i \|^2 \big\}.
  \end{equation}
  Once substituting the proximal regularization in (\ref{argmin_SGD}) by the Kullback–Leibler divergence~\cite{kullback1951information} regularization, it is straightforward to verify that our algorithm is equivalent to the EXP3 algorithm with full information feedback.
  To understand this connection, we can view each agent as a player choosing the desired arm from a bandit while the obtained reward is dynamically affected by all other players. In this sense, our algorithm drives all players to an equilibrium in which nobody can obtain more rewards by unilaterally changing its strategy. Other related works can be found in~\cite{roughgarden2009intrinsic,syrgkanis2015fast,lykouris2016learning,foster2016learning}, which focus on the analysis of regret bound in the context of online learning.
\end{remark}

\section{Distributed Algorithm with\\ Communication Delays}\label{sec:DecentralizedAlgo}

As mentioned before, one major concern of the proposed Algorithm~\ref{algo:SGD} is that individual agents are required to communicate with all others instantaneously, in order to received the sampled strategies $\hat{a}_i^s$'s based on the most updated distributions $\mathbf{p}_i^k$'s. This undoubtedly brings restrictions on the algorithm implementation. In this section, we relax such a requirement and further enhance the proposed algorithm by considering the scenario in which the communication delays are present.

Suppose that each individual agent can only receive others' strategies sampled from the time-delayed distributions. Concretely, let us assume, at the $k$-th iteration, each agent $i$ receives the sampled strategy $\hat{a}_j^s$ from the agent~$j$ which is based on the distribution $\mathbf{p}_j^{k - \tau_{ij}}$. Note that $\tau_{ij}$ here represents the length of time-delays when the agent $i$ receives the information from agent $j$. In addition, to ensure the informational flow between any pair of agents, we restrict, in the following assumption, that the time-delay $\tau_{ij}$ is bounded for any $i,j \in \mathcal{I}$.
\begin{assumption}\label{assump:delayBound}
  It is assumed that there exists a constant $D>0$ such that $\tau_{ij} \le D$ for $\forall i,j \in \mathcal{I}$.
\end{assumption}

We remark that the above Assumption~\ref{assump:delayBound} is quite standard in the study of algorithms with delayed communications. It inherently ensures that each agent receives others' information at least once within the time-window $k \le t \le k+D-1$.

Since the agents' strategies are sampled from the delayed distributions, it is natural to see that the stochastic gradients are also subject to the time-delays. Let us denote the delayed stochastic gradient with respect to $p_i(a_i)$ as
\begin{align}
\nabla_{{p}_i(a_i)} \hat{f}_\delta(P_i^{k-}) =  \frac{1}{M} \cdot \sum_{s=1}^M F(a_i; \hat{{A}}_{- i}^s),
\end{align}
in which we use $P_i^{k-}$ to represent the delayed distributions $\mathbf{p}_j^{k - \tau_{ij}}$'s associated with the agent $i$, and $\hat{{A}}_{- i}^s$ is the set of sampled strategies based on the delayed $P_i^{k-}$. As a result, the iteration of our projected stochastic gradient algorithm with communication delays becomes
\begin{align}\label{delayedSGD}
  \mathbf{p}_i^{k+1} = \Pi_{\mathcal{S}}\big( \mathbf{p}_i^{k} + \gamma \cdot \nabla_{\mathbf{p}_i} \hat{f}_\delta(P_i^{k-}),
\end{align}
where $\nabla_{\mathbf{p}_i} \hat{f}_\delta(P_i^{k-}) $ is the vector that stacks $\nabla_{{p}_i(a_i)} \hat{f}_\delta(P_i^{k-})$'s for all $a_i \in \mathcal{A}_i$.

Herein, let us refer to the scheme~(\ref{delayedSGD}) as our Algorithm~2. As similar to Algorithm~\ref{algo:SGD}, in order to establish the convergence of Algorithm~2, we first characterize the condition of equilibrium solutions when communication delays are present.

\begin{proposition}\label{prop:deGradMapping}
  Suppose that Assumptions~\ref{assump:distinguishable} and \ref{assump:delayBound} hold, and let $\PsD^{k} = [P^k, P^{k+1}, \cdots, P^{k+D-1}]$ be a collection of probability distributions which are generated by Algorithm~2 within the time-window $k\le t\le k+{D}-1$. Then, each $P^t$ within the time-window is an equilibrium solution to problem~(\ref{prob_max}), if the following condition is satisfied,
  \begin{align}\label{deGradMapping}
    \sum_{t = k}^{k+D-1}\mathbb{E} \Big[\big\|\mathbf{p}_i^{t} - \Pi_{\mathcal{S}}\big( \mathbf{p}_i^{t} + \gamma \cdot \nabla_{\mathbf{p}_i} \hat{f}_\delta(P_i^{t-}) \big)\big\|^2\Big] = 0, \; \forall i \in \mathcal{I}.
  \end{align}
\end{proposition}

Now, we establish the convergence of Algorithm~2 by the following theorem.

\begin{theorem}\label{theorem:delay}
    Suppose that Assumptions~\ref{assump:distinguishable} and \ref{assump:delayBound} hold, and let $\{P^k\}_{k \in \mathbb{N}_+}$ be the sequence generated by Algorithm~2 with a small enough constant step-size $\gamma$ and a large enough sample-size $M$. Then, it holds that, for $\forall i \in \mathcal{I}$,
  \begin{align}
  \lim_{k \to \infty} \sum_{t = k}^{k+D-1}\mathbb{E} \Big[\big\|\mathbf{p}_i^{t} - \Pi_{\mathcal{S}}\big( \mathbf{p}_i^{t} + \gamma \cdot \nabla_{\mathbf{p}_i} \hat{f}_\delta(P_i^{t-}) \big)\big\|^2\Big] = 0,
  \end{align}
  and furthermore, the running average converges at the rate of $\mathcal{O}(1/T)$ where $T$ is the number of iterations, i.e., there exists a constant $\eta_2 > 0$ such that
  \begin{align}
    \frac{1}{T}&\sum_{k=0}^T \mathbb{E} \Big[\big\|\mathbf{p}_i^{t} - \Pi_{\mathcal{S}}\big( \mathbf{p}_i^{t} + \gamma \cdot \nabla_{\mathbf{p}_i} \hat{f}_\delta(P_i^{t-}) \big)\big\|^2\Big] \le \frac{\eta_2}{T}.
  \end{align}
\end{theorem} 





As earlier, we present the theoretical proofs of the above Proposition~\ref{prop:deGradMapping} and Theorem~\ref{theorem:delay} in Appendix~D and E, respectively; the detailed conditions of the step-size $\gamma$ and sample-size $M$ are also provided in the proof. In the end of this section, we make a few remarks on the implementation of the enhanced Algorithm~2 with delayed communications.

\begin{remark}
For the first $D$ iterations, individual agents might not be able to receive all others' information, due to the presence of communication delays. Under such circumstance, our algorithm allows the agent to arbitrarily initialize the received strategies, and the convergence of algorithm will not be affected. For instance, each agent can choose the empty as the corresponding strategy if no information is received.  In this sense, we remark that the delayed probability distribution $P_i^{k-}$ is actually well-defined for all $k \ge 0$. 
\end{remark}

\begin{remark}\label{remark:distributedGraph}
  It is also noteworthy that, since the enhanced Algorithm~2 is robust against the communication delays, one can implement it in a fully distributed architecture where agents only need to communicate with their neighbors. Suppose that the communication channels among agents are governed by a peer-to-peer network, denoted as a general graph $G$. Then, the time-delay $\tau_{ij}$ in Algorithm~2 corresponds to the distance~$\delta_{ij}$ between node $i$ and $j$, i.e., the minimum number of edges that connect those two nodes. Therefore, as long as the graph $G$ is connected, meaning that $\delta_{ij}$ is bounded, our algorithm is still effective. In this case, however, the price is that each agent needs to maintain a memory buffer to store the information for all others within the network.
\end{remark}

\section{Simulation}\label{sec:simulation}
In this section, we evaluate the effectiveness of the proposed algorithms by considering a real-world movie recommendation application \cite{ma2019learning,zhang2020explainable}. Our numerical simulations are conducted based on the well-known MovieLens dataset \cite{harper2015movielens}, which contains over $25$ million ratings (ranging from $1$ to~$5$) applied to $62,423$ movies by $162,541$ different users. In particular, we denote $r_{i,j}$ the rating submitted by the user $i$ to the movie $j$, and say that the movie $j$ is liked by the user~$i$ if $r_{i,j} \ge \bar{r}$ where $\bar{r}$ is some certain pre-defined threshold. The objective herein is to identify the top $I$ movies, in the sense that those movies are liked by the maximum number of users. It should be noted that the considered problem is not trivial, since we count each user only once for all the chosen $I$ movies. For example, suppose that the user $i$ likes the movies $j$ and $j'$ at the same time and both of them are chosen as the top movies, then the user $i$ will be counted only once, rather than twice, when counting the number of users for the top movies.

To formalize the above movie recommendation problem, let us denote $\mathcal{S}$ the set of all movies and $\mathcal{U}(j_s): = \{i \, \vert \, r_{i,j_s} \ge \bar{r}\}$ the set of users who like the movie $j_s$. Then, the considered movie recommendation application can be formulated as the following maximization problem,
\begin{align}\label{movieRecommend}
  \max_{j_s \in \mathcal{S}}\quad F(j_1, j_2, \cdots, j_I):=|\cup_{s =1}^I \mathcal{U}(j_s)|.
\end{align}
It can be verified that the objective function $F$ is both monotone and submodular; see definitions in Assumption~\ref{assumption:function}. Thus, (\ref{movieRecommend}) is a well-defined monotone submodular maximization~problem which can be solved by the proposed algorithms.

In our simulations, we specify the rating threshold as $\bar{r} = 3$ and aim to identify the top $I = 10$ movies. In addition, to reduce the size of the candidate movie set~$\mathcal{S}$, we pre-process the dataset and only consider the movies which are liked by no less than 300 users. As a result, totally {1,160} movies are picked up to comprise the candidate set $\mathcal{S}$. In the following, we conduct two separate simulations which are corresponding to the two proposed algorithms. Each simulation is compliant with a network composed of $I=10$ agents. Therefore, each individual agent only needs to take charge in the determination of one (out of {1,160}) movie, so that the entire network cooperatively finds the top 10 movies. In the first simulation, a fully-connected network is assumed, i.e., the all-to-all communications are available for each single agent, so that the Algorithm~\ref{algo:SGD} can be implemented without time-delays. Additionally, in order to take into account the delayed communications, we assume a general but connected undirected network in the second simulation. In this case, agents only communicate with their neighbors. As mentioned in Remark~\ref{remark:distributedGraph}, the length of time-delays $\tau_{ij}$ is governed by the distance between the agent $i$ and $j$ presented in the network.

The following Fig.~\ref{fig:prob_centralized} first plots the evolution of individual agents' probabilities of choosing the final top 10 movies, in the case that the fully-connected network is assumed. Note that the step-size and sample-size are set as $\gamma = 0.0005$ and $M = 3$. As one can observe from Fig.~\ref{fig:prob_centralized}, each agent decides its own choice after around $350$ iterations, and it is confirmed that the collection of the top 10 movies is an equilibrium solution to problem (\ref{movieRecommend}). More specifically, it turns out that the chosen top 10 movies are liked by $10,700$ users, while totally $11,842$ users are involved in all the {1,160} candidate movies. Although it is unknown how many users are covered by the optimal collection of the ten movies, given that this quantity has to be no larger than $11,842$, thus, it is immediately confirmed that the $1/2$-suboptimal bound is achieved.

\begin{figure}
  \centering
  \includegraphics[width=0.9\linewidth]{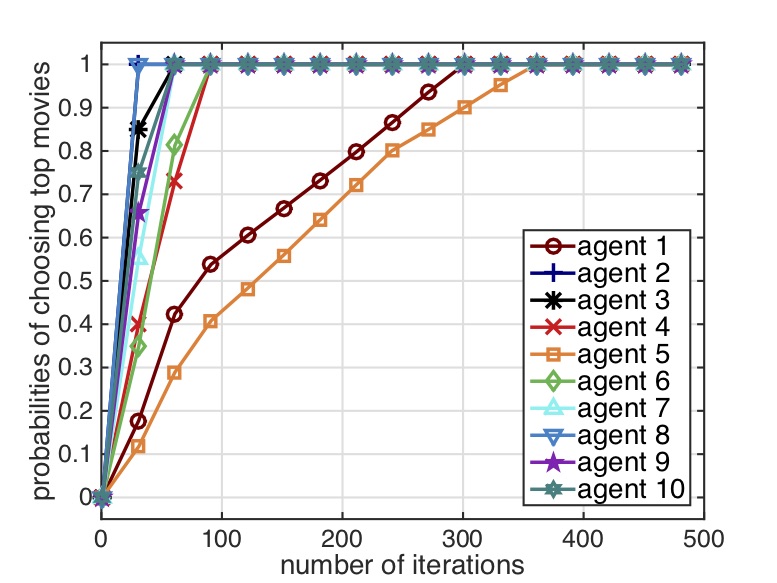}
  \caption{\small Evolution of agents' probabilities (fully-connected graph).}
  \vspace{-10pt}
  \label{fig:prob_centralized}
\end{figure}

\begin{figure}
  \centering
  \includegraphics[width=0.9\linewidth]{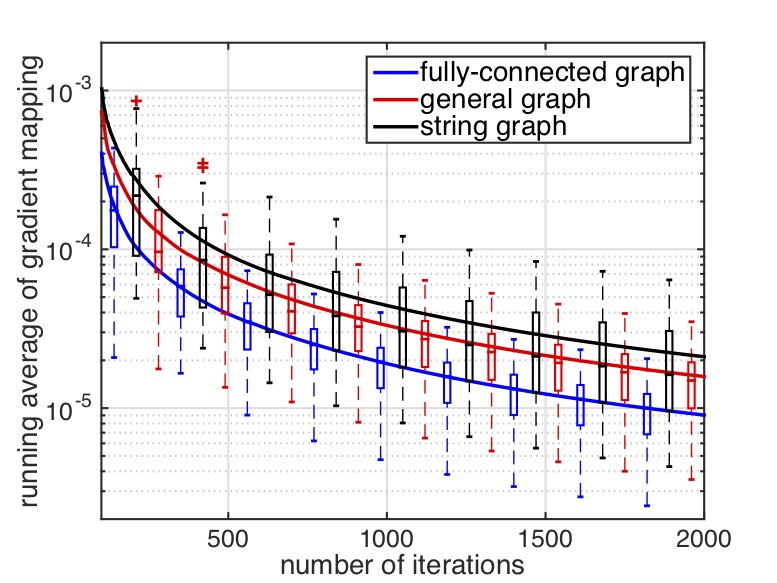}
  \caption{\small Averaged gradient mapping with different graphs.}
  \vspace{-10pt}
  \label{fig:gradient_map}
\end{figure}

Furthermore, in order to evaluate the statistical performance of our stochastic gradient based algorithm, we carry out the Monte-Carlo simulation for $20$ times. The simulation setting is the same as before. Fig.~\ref{fig:gradient_map} plots the running average of the generated gradient mappings (blue curve), averaged by the $20$ independent trials. Note that the running average of gradient mappings $J^k$ at each iteration $k$ is computed as 
\begin{align}
  J^k :=\frac{1}{k}\cdot  \sum_{t=1}^k \sum_{i=1}^I\|\mathbf{p}_i^t - \mathbf{p}_i^{t-1}\|^2.
\end{align}
According to Proposition~\ref{prop:gradMapping}, it is implied that the algorithm converges to the desired equilibrium solution when $J^k \to 0$. Therefore, Fig.~\ref{fig:gradient_map} validates the convergence of the proposed Algorithm~\ref{algo:SGD} and one can also observe a sublinear convergence rate as stated in Theorem~\ref{theorem:SGD}.

\begin{figure}
    \centering
    \begin{subfigure}[b]{0.48\linewidth}
        \centering
        \includegraphics[width=0.9\textwidth]{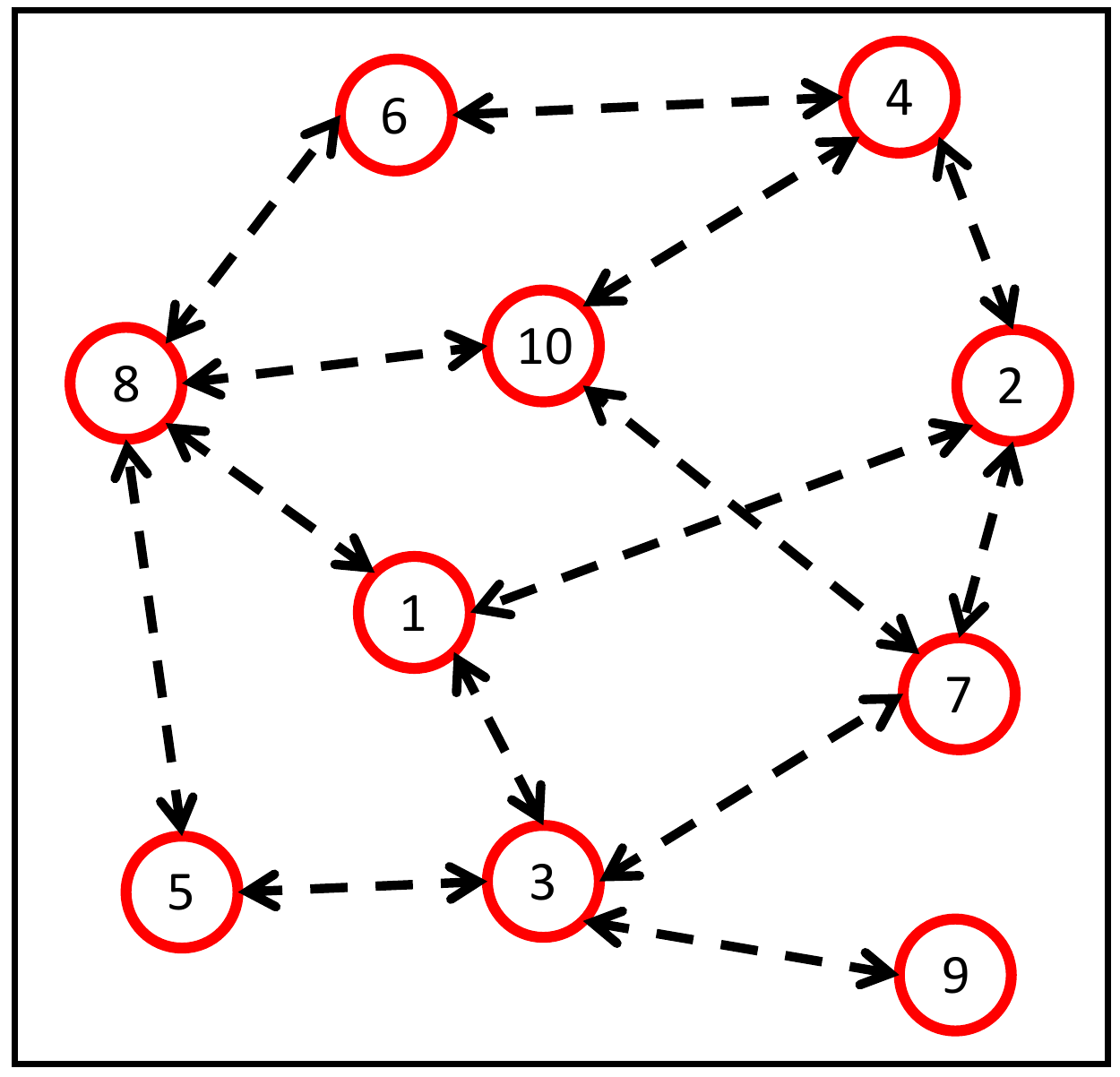}
        \caption[]%
        {{\small The general connected graph}}    
        \label{subfig:network_general}
    \end{subfigure}
    \hfill
    \begin{subfigure}[b]{0.48\linewidth}  
        \centering 
        \includegraphics[width=0.9\textwidth]{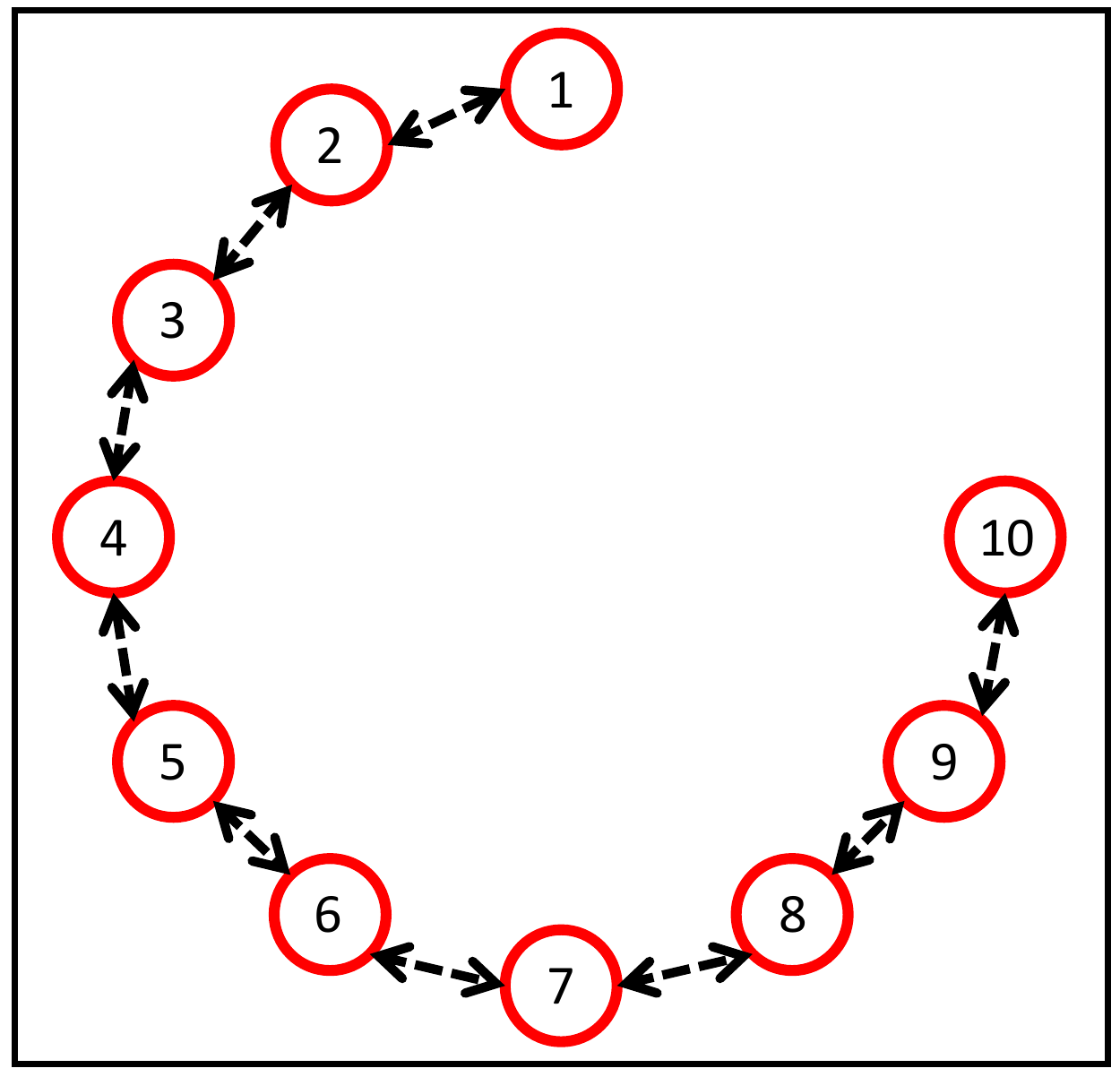}
        \caption[]%
        {{\small The string graph}}    
        \label{subfig:network_string}
    \end{subfigure}
    \caption[]
    {\small Multi-agent network topologies.} 
    \vspace{-10pt}
    \label{fig:topo}
\end{figure}

\begin{figure}
  \centering
  \includegraphics[width=0.9\linewidth]{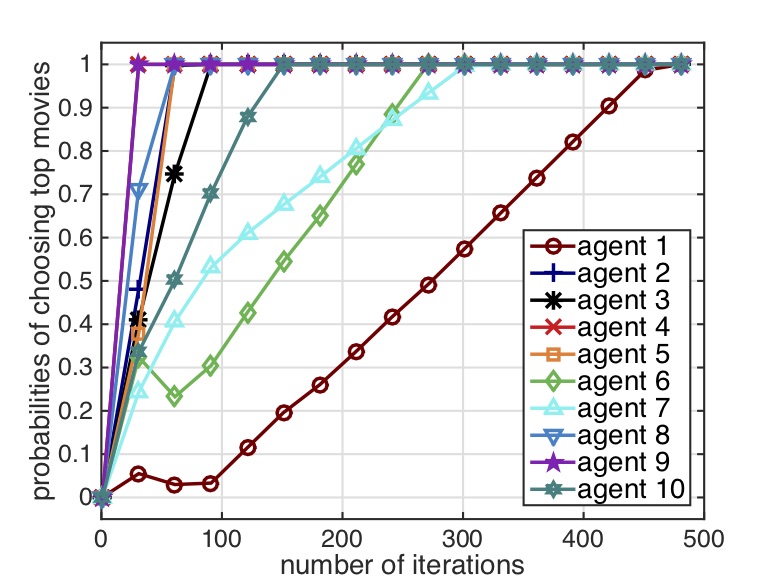}
  \caption{\small Evolution of agents' probabilities (general graph).}
  \vspace{-10pt}
  \label{fig:prob_network}
\end{figure}

In the second simulation, we run the proposed Algorithm~2 with delayed communications, under a general connected network as shown in Fig.~\ref{subfig:network_general}. It can be seen that from the topology that the maximum distance between two nodes is four, and thus the communication delays are bounded by the constant $D = 3$. Fig.~\ref{fig:prob_network} shows the the evolution of agents' probabilities of choosing the top 10 movies, in the case that $\gamma = 0.0005$ and $M = 3$. Based on this figure, we conclude that Algorithm~2 also converges with the presence of delayed communications. However, compared to the case without time-delays as shown in Fig.~\ref{fig:prob_centralized}, more iterations are needed to arrive at the final decisions of the top 10 movies. 

Moreover, as similar to the first simulation, we conduct the Monte-Carlo simulation with $20$ independent trails as well. Besides the general connected network as shown in Fig.~\ref{subfig:network_general}, in this simulation, we additionally consider a specific string graph as shown in Fig.~\ref{subfig:network_string}. It should be noted that, under such a string graph, the maximum distance between two nodes is nine and thus the communication delays are bounded by $D = 8$. The averaged $J^k$'s in these two cases are also demonstrated in Fig.~\ref{fig:gradient_map}; see the red curve for the general connected graph and the black curve for the string graph. Based on Proposition~\ref{prop:deGradMapping}, it is confirmed that Algorithm 2 converges to the equilibrium solution and a sublinear convergence rate is shown as expected. In addition, we also observe from the figure that a larger number of iterations are needed to obtain the solution when more communication delays are present in the network.

\section{Conclusion}\label{sec:conclusion}
{In this paper, 
we developed a projected stochastic gradient algorithm for solving the distributed submodular maximization problem. Unlike the commonly-studied greedy algorithm, our approach enables the simultaneous updates among all individual agents. It is proved that the algorithm converges to an equilibrium solution, which is guaranteed to be at least $1/2$-suboptimal. 
Furthermore, we enhanced the proposed algorithm by handling the scenario in which agents’ communication delays are present. The similar convergence result is proved for the enhanced distributed algorithm. Finally, a real-world movie recommendation application is considered to demonstrate the effectiveness of our algorithms.}
{It should be also remarked that, compared to the brute-force searching scheme which has exponential complexity in terms of the number of agents $I$, the complexity of our distributed algorithms is primarily reduced.
We leave the rigorous complexity analysis of the algorithms as our future work.}

\section*{Appendix}

\subsection{Proof of Proposition~\ref{prop:suboptimal}}

According to the definition of the function $F(A)$, let us first define its marginal function as
\begin{align}
  \delta(a \,\vert \,A) := F(\{a\} \cup A) - F(A),
\end{align}
It can be shown that the submodularity of function $F(A)$ implies that if $A' \subseteq A$, then
\begin{align}
  \delta(a \,\vert \,A' ) \ge   \delta(a \,\vert \,A).
\end{align}
Recall that we use $A_{-i}$ to denote the set of elements $a_j$'s in $A$ where \mbox{$j \neq i$}. In addition, we use $A_{<i}$ (or $A_{\le i}$) to represent the set of elements $a_j$  where $j <i$ (or $j \le i$). Then, by the definition of the equilibrium $A^e$; see Definition~\ref{def:equilibrium}, we can have
\begin{align}\label{eqMarginal}
  \delta(a_i^e\, \vert \,A^e_{-i}) \ge \delta(a_i\, \vert \,A^e_{-i}), \; \forall a_i \in \mathcal{A}_i, i \in \mathcal{I}.
\end{align}

Now, based on Assumption~\ref{assumption:function} of the function $F(A)$ (monotonicity and submodularity), it holds that
\begin{equation}
  \begin{aligned}
    F(A^\star) &\overset{(1a)}{\le} F(A^\star \cup A^e)\\
    & = F(A^e)+ \sum_{i=1}^I \Big(F(A^\star_{\le i} \cup A^e) - F(A^\star_{< i} \cup A^e) \Big)\\
    & \overset{(1b)}{=}  F(A^e) + \sum_{i=1}^I \delta(a_i^\star \,\vert\,A^\star_{< i} \cup A^e)\\
    & \overset{(1c)}{\le} F(A^e) + \sum_{i=1}^I \delta(a_i^\star \,\vert\,A_{-i}^e)\\
    & \overset{(1d)}{\le} F(A^e) + \sum_{i=1}^I \delta(a_i^e \,\vert\,A_{-i}^e)\\
    & \overset{(1e)}{\le} F(A^e) + \sum_{i=1}^I \delta(a_i^e \,\vert\,A_{<i}^e)\\
    & \overset{(1f)}{=} F(A^e) + \sum_{i=1}^I \Big(F( A_{\le i}^e) - F(A^e_{< i}) \Big)\\
    & \overset{(1g)}{=} 2F(A^e).
  \end{aligned}
\end{equation}
Note that $(1a)$ comes from the monotonicity of $F(A)$; $(1b)$ and $(1f)$ is due to the definition of marginal function; $(1c)$ and $(1e)$ comes from the submodularity of $F(A)$; $(1d)$ is based on the inequality (\ref{eqMarginal}); and $(1g)$ is due to the fact that $F(\emptyset) = 0$. Therefore, the proof is completed.

\subsection{Proof of Proposition \ref{prop:gradMapping}}
 We start the proof by investigating properties of the projection on the probability simplex $\mathcal{S}$. According to the definition of the projection $\Pi_\mathcal{S}(\cdot)$, as shown in~(\ref{projection}), it can be verified in \cite{duchi2008efficient,wang2013projection} that the projection is computed as 
\begin{align}\label{plus}
  \Pi_{\mathcal{S}}(\mathbf{p}) = [\mathbf{p} - \lambda\mathbf{1}]^+,
\end{align}
where $\lambda$ is the solution of the equation $\mathbf{1}^\top[\mathbf{p} - \lambda\mathbf{1}]^+ = 1$. Subsequently, we show the following two lemmas regarding the projection $\Pi_{\mathcal{S}}(\mathbf{p})$.

\begin{lemma}\label{lemma:1}
    Suppose that $\mathbf{p} \in [0,1]^K$ is a vertex of the simplex, i.e., $\|\mathbf{p}\|_\infty = 1$, and let us denote its non-zero component as $\mathbf{p}(n)$, i.e., $\mathbf{p}(n) = 1$ and $\mathbf{p}(k) = 0$ for $\forall k \neq n$. Then, one can have $ \mathbf{p} = \Pi_{\mathcal{S}} (\mathbf{p} + \Delta_{\mathbf{p}}) $ if and only if $\Delta_{\mathbf{p}} (n)$ is the maximum component of $\Delta_{\mathbf{p}}$, i.e., $\Delta_{\mathbf{p}} (n) \ge \Delta_{\mathbf{p}} (k)$ for $\forall k = 1,2,\cdots,K$.
\end{lemma}

\begin{IEEEproof}
  Let us recall that $\Pi_{\mathcal{S}} (\mathbf{p} + \Delta_{\mathbf{p}}) = [\mathbf{p} + \Delta_{\mathbf{p}} - \lambda \mathbf{1}]^+$. Suppose that $ \mathbf{p} = \Pi_{\mathcal{S}} (\mathbf{p} + \Delta_{\mathbf{p}}) $, since $\mathbf{p}$ is a vertex of the simplex and $\mathbf{p}(n) = 1$, then the $n$-th component of vector $\mathbf{p} + \Delta_{\mathbf{p}} - \lambda \mathbf{1}$ must be one, and all its other components must be non-positive. It means that $\mathbf{p}(n) + \Delta_{\mathbf{p}}(n) - \lambda = 1$ and $\mathbf{p}(k) + \Delta_{\mathbf{p}}(k) - \lambda \le 0$ for $\forall k \neq n$. The former equality tells that $\Delta_{\mathbf{p}}(n) =  \lambda$ and the latter yields $\Delta_{\mathbf{p}}(k) \le  \lambda$. Combining those two proves the first half of the statement.

   Conversely, assume $\Delta_{\mathbf{p}} (n) \ge \Delta_{\mathbf{p}} (k)$ for $\forall k = 1,2,\cdots,K$, then it holds that $\big(\mathbf{p}(n) + \Delta_{\mathbf{p}}(n)\big) - \big(\mathbf{p}(k)+ \Delta_{\mathbf{p}}(k)\big) \ge 1$. Thus, to satisfy the equation $\mathbf{1}^\top[\mathbf{p} + \Delta_{\mathbf{p}} - \lambda \mathbf{1}]^+ = 1$, we must have $\lambda = \Delta_{\mathbf{p}}(n)$. On this account, we can further have that $\mathbf{p}(n) + \Delta_{\mathbf{p}}(n) - \lambda = 1$ and $\mathbf{p}(k) + \Delta_{\mathbf{p}}(k) - \lambda \le 0$ for $\forall k \neq n$, and thus $[\mathbf{p} - \Delta_{\mathbf{p}} - \lambda \mathbf{1}]^+ = \mathbf{p}$. Therefore, the second part of the statement is proved.
\end{IEEEproof}

\begin{lemma}\label{lemma:2}
  Suppose that $\mathbf{p}\in [0,1]^K$ is not a vertex of the simplex, and without loss of generality, let us assume its first $n$ components ($2 \le n \le K$) to be non-zeros. Then, one can have $ \mathbf{p} = \Pi_{\mathcal{S}} (\mathbf{p} + \Delta_{\mathbf{p}}) $ if and only if there exists $\delta$ such that $\Delta_{\mathbf{p}}(k) = \delta $ for $k \le n$ and $\Delta_{\mathbf{p}}(k) \le \delta $ for $k > n$.
\end{lemma}
\begin{IEEEproof}
  Recall again that  $\Pi_{\mathcal{S}} (\mathbf{p} + \Delta_{\mathbf{p}}) = [\mathbf{p} + \Delta_{\mathbf{p}} - \lambda \mathbf{1}]^+$. Let us assume $ \mathbf{p} = \Pi_{\mathcal{S}} (\mathbf{p} + \Delta_{\mathbf{p}}) $, since the first $n$ components of $\mathbf{p}$ are non-zeros, then we have $\mathbf{p}(k) = \mathbf{p}(k) + \Delta_{\mathbf{p}}(k) - \lambda$ for $k \le n$ and $ \Delta_{\mathbf{p}}(k) -\lambda \le 0 $ for $k > n$. Thus, simply taking $\delta = \lambda$ proves the first half of the statement. 

  Conversely, suppose that $\Delta_\mathbf{p}$ has $\Delta_{\mathbf{p}}(k) = \delta $ for $k \le n$ and $\Delta_{\mathbf{p}}(k) \le \delta $ for $k > n$. Since it is known that $\mathbf{1}^\top \mathbf{p} = 1$, in order to ensure the equation $\mathbf{1}^\top[\mathbf{p}  + \Delta_\mathbf{p} - \lambda\mathbf{1}]^+ = 1$, we must have $\delta = \lambda$. Thus, it holds that $ \Pi_{\mathcal{S}} (\mathbf{p} + \Delta_{\mathbf{p}}) = [\mathbf{p}]^+ = \mathbf{p}$.
\end{IEEEproof}

With the help of the above two lemmas, we are now ready to prove the proposition in both directions separately.

\vspace{5pt}
\noindent{\textbf{Definition}} \ref{def:Pe} $\Rightarrow$ {\textbf{Equation}} (\ref{gradMapping}):
\vspace{5pt}

Let us assume that $P^e = [{\mathbf{p}^e_1}^\top, {\mathbf{p}^e_2}^\top, \cdots\hspace{-2pt}, {\mathbf{p}^e_I}^\top]^\top$ is an equilibrium following Definition~\ref{def:Pe}. According to the definition, we know that each $\mathbf{p}_i^e$ must be a vertex of the simplex $\mathcal{S}$, i.e., there exists $a_i^e$ such that $p_i^e(a_i^e) = 1$ and $p_i^e(a_i) = 0$ for $\forall a_i \neq a_i^e$; and in addition, the collection of $a_i^e$'s, i.e., $A^e = [a_1^e,a_2^e, \cdots,a_I^e]^\top$, is the equilibrium following Definition~\ref{def:equilibrium}. Provided that $P^e$ is the collection of simplex vertices and the stochastic gradient $\nabla_{\mathbf{p}_i} \hat{f}({{P}}^e)$ in (\ref{gradMapping}) is sampled based on the probability distributions $\mathbf{p}_i^e$'s, thus it can be shown that the stochastic gradient has the following fixed form
\begin{align}
  \nabla_{\mathbf{p}_i} \hat{f}({{P}}^e) = [F(a_i; A^e_{-i})]_{a_i \in \mathcal{A}_i},
\end{align}
and we can simply get rid of the expectation in (\ref{gradMapping}). Moreover, by the definition of equilibrium $A^e$ (see equation (\ref{eq:equilibrium})), it holds that, for $\forall i \in \mathcal{I}$,
\begin{align}
  F(a_i^e; A^e_{-i}) \ge F(a_i; A^e_{-i}),\; \forall a_i \in \mathcal{A}_i.
\end{align}
On this account,  we know that each gradient  $\nabla_{\mathbf{p}_i} \hat{f}({{P}}^e)$ has $F(a_i^e; A^e_{-i})$ as its maximum component. Therefore, based on Lemma~\ref{lemma:1}, it is proved that
\begin{align}
  \mathbf{p}_i^e = \Pi_{\mathcal{S}}\big( \mathbf{p}_i^{e} +\gamma \cdot \nabla_{\mathbf{p}_i} \hat{f}({{P}}^e) \big),
\end{align}
and thus the proof of the first half is completed.

\vspace{5pt}
\noindent{{\textbf{Equation}} (\ref{gradMapping}) $\Rightarrow$ \textbf{Definition}} \ref{def:Pe}:
\vspace{5pt}

Suppose that the point $P^e$ has already satisfied the condition~(\ref{gradMapping}). Next, we first show that each $\mathbf{p}_i^e$ in $P^e$ has to be the vertex of the simplex $\mathcal{S}$. In fact, suppose that $\mathbf{p}_i^e$ is not a vertex and let $p_i^e(a_i^1)$ and $ p_i^e(a_i^2)$ be the two non-zero components. Then, based on Lemma~\ref{lemma:2} and in order to ensure the condition~(\ref{gradMapping}), we must have that for $\forall a_i \in \mathcal{A}_i$,
\begin{align}
 \sum_{s=1}^M F(a_i^1; \hat{{A}}_{- i}^s)= \sum_{s=1}^M F(a_i^2; \hat{{A}}_{- i}^s) {\ge \sum_{s=1}^M F(a_i; \hat{{A}}_{- i}^s)},
\end{align}
where $\hat{A}_{-i}^s$ represents any possible sample of strategies. This clearly contradicts the {maximum distinguishable assumption} of the function $F(A)$ (see Assumption~\ref{assump:distinguishable}). As a result, we have proved that $P^e$ must be the collection of simplex vertices, and the expectation in~(\ref{gradMapping}) can be removed.
Next, let us assume that each $\mathbf{p}_i^e$ has $p_i^e(a_i^e) = 1$ and $p_i^e(a_i) = 0$ for $\forall a_i \neq a_i^e$, by the fact that it is simply a vertex. Applying Lemma~\ref{lemma:1} once again, it can be shown that the gradient $\nabla_{\mathbf{p}_i} \hat{f}({{P}}^e)$ has its maximum component at $F(a_i^e; A^e_{-i})$, i.e.,
\begin{align}
  a_i^e = \argmax_{a_i \in \mathcal{A}_i} \; F(a_i; A^e_{-i}).
\end{align}
Thus, the second half of the proposition is proved.

\subsection{Proof of Theorem~\ref{theorem:SGD}}

We begin the proof by recalling that the stochastic gradient $\nabla_{\mathbf{p}_i} \hat{f}(P^k)$ computed as (\ref{stoGrad}) is an unbiased estimation of the full gradient $\nabla_{\mathbf{p}_i} {f}(P^k)$. In fact, this statement has been verified in the derivation of our projected stochastic gradient algorithm; see Section~\ref{subsec:algorithm}. Thus, to facilitate the subsequent proof, we here extract the statement as the following lemma and omit the detailed proof.
\begin{lemma}\label{lemma:unbiasedGrad}
  Suppose that the stochastic gradient $\nabla_{\mathbf{p}_i} \hat{f}(P)$ is defined as (\ref{stoGrad}), then it holds that
  \begin{align}
    \mathbb{E}_{\hat{a}_j \sim \mathbf{p}_j,j \neq i}\big[\nabla_{\mathbf{p}_i} \hat{f}(P)\big] = \nabla_{\mathbf{p}_i} {f}(P).
  \end{align}
\end{lemma}

Next, let us introduce an additional notion, namely \textit{gradient mapping}, which is defined as below,
\begin{align}
   \mathcal{G}_\gamma(\mathbf{g}, \mathbf{p}) = \frac{1}{\gamma} \cdot \big(\mathbf{p} - \Pi_\mathcal{S}(\mathbf{p} + \gamma \mathbf{g})\big).
\end{align}
Here, $\mathbf{g} \in \mathbb{R}^K$ a general gradient, $\mathbf{p} \in \mathcal{S}$ is a general point from the probability simplex, and $\gamma$ is a constant which represents the step-size. Recall that our projected stochastic gradient algorithm performs the iteration~(\ref{SGD}), it can be equivalently rewritten into the following gradient mapping form,
\begin{align}\label{lineSearch}
  \mathbf{p}_i^{k+1} = \mathbf{p}_i^k - \gamma\cdot \mathcal{G}_\gamma(\nabla_{\mathbf{p}_i} \hat{f}(P^k), \mathbf{p}_i^k).
\end{align}
The iteration (\ref{lineSearch}) can be interpreted as a standard line search algorithm with the constant step-size $\gamma$, while the searching direction is the gradient mapping $\mathcal{G}_\gamma(\nabla_{\mathbf{p}_i} \hat{f}(P^k), \mathbf{p}_i^k)$. 

Associated with the gradient mapping, we next show the following two lemmas, which will play key roles in the proof of the theorem.

\begin{lemma}\label{lemma:GradMapping1}
  Given the gradient mapping $\mathcal{G}_\gamma(\mathbf{g}, \mathbf{p})$, for any $\mathbf{g} \in \mathbb{R}^K$, $\mathbf{p} \in \mathcal{S}$ and $\gamma \in \mathbb{R}_+$, it holds that
  \begin{align}
    -\langle \mathbf{g}, \; \mathcal{G}_\gamma(\mathbf{g}, \mathbf{p})\rangle \ge \|\mathcal{G}_\gamma(\mathbf{g}, \mathbf{p})\|^2.
  \end{align}
\end{lemma}
\begin{IEEEproof}
  Recall that the projection on the probability simplex $\Pi_\mathcal{S} (\cdot)$ is defined as (\ref{projection}). Thus, within the gradient mapping, the term $\Pi_\mathcal{S}(\mathbf{p} + \gamma \mathbf{g})$ can be computed as
  \begin{align}\label{Proj_gradMapping}
    \Pi_\mathcal{S}(\mathbf{p} + \gamma \mathbf{g}) = \argmin_{\mathbf{x} \in \mathcal{S}}\; \{-\langle \mathbf{g},\; \mathbf{x} \rangle + \frac{1}{2\gamma}\|\mathbf{x} - \mathbf{p}\|^2\}.
  \end{align}
  By noticing that the optimization problem in (\ref{Proj_gradMapping}) is convex, the optimality condition of solution $\Pi_\mathcal{S}(\mathbf{p} + \gamma \mathbf{g})$ ensures that, for $\forall \mathbf{x} \in \mathcal{S}$,
  \begin{align}\label{optCondition}
    \big\langle -\mathbf{g} + \frac{1}{\gamma}\big(\Pi_\mathcal{S}(\mathbf{p} + \gamma \mathbf{g}) - \mathbf{p}\big),\; \mathbf{x} - \Pi_\mathcal{S}(\mathbf{p} + \gamma \mathbf{g})\big\rangle \ge 0.
  \end{align}
  Now, let $\mathbf{x} = \mathbf{p}$, it can be shown that
  \begin{align}
    -\big\langle \mathbf{g},\; \mathbf{p} - \Pi_\mathcal{S}(\mathbf{p} + \gamma \mathbf{g})\big \rangle \ge \frac{1}{\gamma}\|\mathbf{p} - \Pi_\mathcal{S}(\mathbf{p} + \gamma \mathbf{g})\|^2.
  \end{align}
  Provided that the gradient mapping $\mathcal{G}_\gamma(\mathbf{g}, \mathbf{p})$ is define as (\ref{gradMapping}), thus the proof is completed.
\end{IEEEproof}

\begin{lemma}\label{lemma:GradMapping2}
   Given the gradient mapping $\mathcal{G}_\gamma(\mathbf{g}, \mathbf{p})$, for any $\mathbf{g}_1, \mathbf{g}_2 \in \mathbb{R}^K$, $\mathbf{p} \in \mathcal{S}$ and $\gamma \in \mathbb{R}_+$, it holds that
   \begin{align}
     \|\mathcal{G}_\gamma(\mathbf{g}_1, \mathbf{p}) - \mathcal{G}_\gamma(\mathbf{g}_2, \mathbf{p})\| \le \|\mathbf{g}_1 - \mathbf{g}_2\|.
   \end{align}
   \begin{IEEEproof}
     Applying again the optimality condition (\ref{optCondition}) with the gradient $\mathbf{g}$ substituted by $\mathbf{g}_1$ and $\mathbf{g}_2$ respectively, it yields that,
     \begin{subequations}
       \begin{align}
          \big\langle \hspace{-3pt}-\mathbf{g}_1\hspace{-2pt} + \hspace{-2pt}\frac{1}{\gamma}\big(\Pi_\mathcal{S}(\mathbf{p} \hspace{-1pt}+\hspace{-1pt} \gamma \mathbf{g}_1)\hspace{-2pt} - \hspace{-2pt}\mathbf{p}\big),\, \mathbf{x} \hspace{-2pt}-\hspace{-2pt} \Pi_\mathcal{S}(\mathbf{p} + \gamma \mathbf{g}_1)\big\rangle \ge 0;\label{subeq1}\\
          \big\langle \hspace{-3pt}-\mathbf{g}_2\hspace{-2pt} + \hspace{-2pt}\frac{1}{\gamma}\big(\Pi_\mathcal{S}(\mathbf{p} \hspace{-1pt}+\hspace{-1pt} \gamma \mathbf{g}_2)\hspace{-2pt} - \hspace{-2pt}\mathbf{p}\big),\, \mathbf{x} \hspace{-2pt}-\hspace{-2pt} \Pi_\mathcal{S}(\mathbf{p} + \gamma \mathbf{g}_2)\big\rangle \ge 0.\label{subeq2}
       \end{align}
     \end{subequations}
      Now, let $\mathbf{x} = \Pi_\mathcal{S}(\mathbf{p} + \gamma \mathbf{g}_2)$ in (\ref{subeq1}) and $\mathbf{x} = \Pi_\mathcal{S}(\mathbf{p} + \gamma \mathbf{g}_1)$ in~(\ref{subeq2}), summing both inequalities gives that
      \begin{equation}
	      \begin{aligned}
        &\big\langle \mathbf{g}_2 - \mathbf{g}_1 , \; \Pi_\mathcal{S}(\mathbf{p} + \gamma \mathbf{g}_2) -  \Pi_\mathcal{S}(\mathbf{p} + \gamma \mathbf{g}_1)\big\rangle \\
        & \ge \frac{1}{\gamma}\|\Pi_\mathcal{S}(\mathbf{p} + \gamma \mathbf{g}_1) -  \Pi_\mathcal{S}(\mathbf{p} + \gamma \mathbf{g}_2)\|^2
      \end{aligned}
      \end{equation}

      By the definition of gradient mapping $\mathcal{G}_\gamma(\mathbf{g}, \mathbf{p})$ and Cauchy-Schwartz inequality, the proof is completed.
   \end{IEEEproof}
\end{lemma}

It should be remarked that, while Lemma \ref{lemma:unbiasedGrad} verifies that the stochastic gradient $\nabla_{\mathbf{p}_i} \hat{f}(P^k)$ is an unbiased estimation, Lemma \ref{lemma:GradMapping1} and $\ref{lemma:GradMapping2}$ both characterize the properties of the gradient mapping. Next, let us show another lemma which investigates the variance of the stochastic gradient. Before stating the lemma, some more notations and a supporting lemma are needed to be first introduced. Let us simply denote $\nabla {f}(P)$~(also $\nabla \hat{f}(P)$) the stacked full (stochastic) gradient for each agent $i \in \mathcal{I}$, i.e., $\nabla f(P) = [\nabla_{\mathbf{p}_i} {f}(P)]_{i \in \mathcal{I}}$. Similarly, we use $\tilde{\mathcal{G}}_\gamma\big(\nabla f(P),P\big)$ and $\tilde{\Pi}_{\mathcal{S}}\big( P + \gamma\cdot\nabla \hat{f}({{P}}) \big)$ to denote the stacked gradient mapping and also the updated probability distributions respectively, i.e.,
\begin{equation}\label{compactAlgo}
  \begin{aligned}
  P^{k+1}& = \tilde{\Pi}_{\mathcal{S}}\big( P^{k} + \gamma\cdot\nabla \hat{f}({{P}}^k) \big)\\
  & = P^k - \gamma\cdot\tilde{\mathcal{G}}_\gamma\big(\nabla \hat{f}(P^k),P^k\big).
\end{aligned}
\end{equation}

\begin{lemma}\label{lemma:vertices}
  Suppose that the current iterate $P^k$ is a collection of simplex vertices but not an equilibrium. Let the step-size satisfy $\gamma < 2/\Delta^\text{max}$, then the next iterate $P^{k+1}$ generated by Algorithm~\ref{algo:SGD} must not be the collection of simplex vertices.
\end{lemma}

\begin{IEEEproof}
  Let us first recall that the iterate $P^k$ is a collection of $\mathbf{p}_i^k$'s, i.e., $P^k = [{\mathbf{p}^k_1}^\top, {\mathbf{p}^k_2}^\top, \cdots, {\mathbf{p}^k_I}^\top]^\top$. Since it is assumed that each $\mathbf{p}_i^k$ is the simplex vertex, then we let $\mathbf{p}_i^k = \mathbf{e}_{n_i}$ with $\mathbf{e}_{n_i} \in \mathbb{R}^K$ being the unit vector whose $n_i$-th component is one and others are zeros. In addition, according to the iteration of Algorithm~\ref{algo:SGD} and the fact that the projection $\Pi_\mathcal{S}$ can be computed as (\ref{plus}), thus we know
  \begin{align}\label{compP}
    \mathbf{p}_i^{k+1} = \big[\mathbf{e}_{n_i} + \gamma\cdot F_{-i} -\lambda\mathbf{1}\big]^+,
  \end{align}
  where $\lambda$ is governed by the equation $\mathbf{1}^\top\mathbf{p}_i^{k+1} = 1$. Note that here the sampled gradient is deterministic since $P^k$ is a collection of vertices, thus we use $F_{-i} \in \mathbb{R}^K$ to represent the sampled gradient based on the probability distribution $\mathbf{p}_j^k,\, j\neq i$. Furthermore, we denote $F_{-i}(n)$ the $n$-th component of the vector $F_{-i}$.

  Given that $P^k$ is not an equilibrium, thus there exist indices $i \in \mathcal{I}$ and $n_i^\text{max} \neq n_i$, such that $F_{-i}(n_i^\text{max}) > F_{-i}(n_i)$ and
  \begin{align}
    F_{-i}(n_i^\text{max}) \ge F_{-i}(n),\; \forall n = 1,2,\cdots,K.
  \end{align}
  In fact, if $F_{-i}(n_i)$'s are the maximum components for $\forall i \in \mathcal{I}$, then $P^k$ must be the equilibrium by definition. Consequently, according to the equation~(\ref{compP}), we know that $\mathbf{p}_i^{k+1}(n_i^\text{max}) >0$ must be true. On this basis, in order to prove the lemma, it will suffice to show that $\mathbf{p}_i^{k+1}(n_i^\text{max}) < 1$ if $\gamma < 2/\Delta^\text{max}$. Next, we prove this statement by contradiction.

  Suppose that $\mathbf{p}_i^{k+1}(n_i^\text{max}) < 1$ is false, i.e., $\mathbf{p}_i^{k+1}(n_i^\text{max}) = 1$. Provided that $n_i^\text{max} \neq n_i$, thus we have $\gamma \cdot F_{-i}(n_i^\text{max}) - \lambda = 1$ and $1 + \gamma\cdot F_{-i}(n_i) - \lambda \le 0$. Substitute the former equation to the latter one and get rid of $\lambda$, it yields,
  \begin{align}\label{gamma1}
    2 + \gamma\cdot\big(F_{-i}(n_i) - F_{-i}(n_i^\text{max})\big) \le 0.
  \end{align}
  Recall {the definition of $\Delta^\text{max}$; see equation (\ref{deltaMax}), and the fact that $F_{-i}(n_i^\text{max}) > F_{-i}(n_i)$}, we have 
  \begin{align}\label{gamma2}
     0< F_{-i}(n_i^\text{max}) -F_{-i}(n_i) \le \Delta^\text{max}.
  \end{align}
  Combining both (\ref{gamma1}) and (\ref{gamma2}) shows that $\gamma \ge 2/\Delta^\text{max}$. Thus, the proof is completed.

\end{IEEEproof}

Now, we are ready to show the following lemma which characterizes the variance of stochastic gradients.
\begin{lemma}\label{lemma:boundedVar}
Suppose that the sequence $\{P^k\}_{k \in \mathbb{N}_+}$ is the set of iterates generated by Algorithm~\ref{algo:SGD} and the initialization $P^0$ is not a collection of simplex vertices. {Let the step-size~$\gamma$ satisfy the condition $\gamma < 2/\Delta^\text{max}$}, then there exist constants $B_0 >0$ and $ B_1 >0$ such that the following holds,
\begin{equation}\label{var}
  \begin{aligned}
  \sum_{k = 0}^T\mathbb{E}\Big[\|\nabla \hat{f}(P^k) &- \nabla {f}(P^k)\|^2 \Big]\le B_0 \\[-5pt]
  &+ B_1/M \cdot\sum_{k=0}^T\mathbb{E}\Big[\|\tilde{\mathcal{G}}_\gamma\big(\nabla \hat{f}(P^k), P^k\big)\|^2\Big],
\end{aligned}
\end{equation}
where the expectation is taken with respect to the sampling of stochastic gradient $\nabla \hat{f}(P)$ for all iterations $0 \le k \le T$ and $M$ is the sample-size.
\end{lemma}

\begin{IEEEproof}
Before starting the proof, we first note that it is only needed to consider the case when none of~\mbox{$P^k, \; 0 \le k \le T$} is the equilibrium. In fact, it can be immediately verified that the algorithm will stay at the equilibrium $P^e$ forever once it reaches the point. In addition, due to the fact that
\begin{align}
   \mathbb{E}\big[\|\nabla \hat{f}(P^e) &- \nabla {f}(P^e)\|^2 \big] = \mathbb{E}\big[\|\tilde{\mathcal{G}}_\gamma\big(\nabla \hat{f}(P^e), P^e\big)\|^2\big] = 0,
 \end{align} 
 thus we only need to prove the case in which the algorithm has not reach the equilibrium.

Now, let us begin the proof by introducing an additional notion, namely the reachable set of the iterates $P^k$. We define the reachable set $\mathcal{S}^k$ at each iteration $k$ as follows,
  \begin{align}
    \mathcal{S}^k:= \big\{P^k \,|\, P^t = \tilde{\Pi}_{\mathcal{S}}\big( P^{t-1} + \gamma\cdot\nabla \bar{f}({{P}}^{t-1}) \big), 1 \le t \le k\big\}.
  \end{align}
 Note that here the gradient $\nabla \bar{f}({{P}}^{t-1})$ is any possible realization of the stochastic gradient $\nabla \hat{f}({{P}}^{t-1})$. Due to the fact that each  $\nabla \bar{f}({{P}}^{t-1})$ only has finite possibilities and $P^0$ is well initialized, thus we know each $\mathcal{S}^k$ is also a finite set, but its cardinality grows quickly as the index $k$ increases. Subsequently, let us divide each of the reachable sets $\mathcal{S}^k$ into two subsets, i.e., $\mathcal{S}^k = \bar{\mathcal{S}}^k \cup \bar{\mathcal{S}}_c^k$ where $\bar{\mathcal{S}}^k$ only contains the iterates $P^k$'s which are collections of the simplex vertices and $\bar{\mathcal{S}}_c^k$ is the complement set. On this account, we next prove the following statements: there exists a constant $\epsilon > 0$ such that
\begin{enumerate}
    \item if it is known that $P^{k+1} \in \bar{\mathcal{S}}_c^{k+1}$, then
  \begin{equation}\label{statement2}
      \begin{aligned}
\mathbb{E}\big[\|\tilde{\mathcal{G}}_\gamma\big(\nabla \hat{f}(P^k), P^k\big)\|^2\big] \ge \epsilon;
  \end{aligned}
  \end{equation}

    \item if it is known that $P^{k} \in \bar{\mathcal{S}}^{k}$, then
  \begin{equation}\label{statement1}
      \begin{aligned}
\mathbb{E}\big[\|\tilde{\mathcal{G}}_\gamma\big(\nabla \hat{f}(P^k), P^k\big)\|^2\big] \ge \epsilon.
  \end{aligned}
  \end{equation}
\end{enumerate}

\textit{Proof of statement 1)}: Recall again that the iterate $P^k$ is the collection of $\mathbf{p}_i^k$'s. Since it is known that $P^{k+1} \in \bar{\mathcal{S}}_c^{k+1}$, then let us assume, without loss of generality, that $\mathbf{p}_i^{k+1}$ has two non-zero components $\mathbf{p}_i^{k+1} (u)$ and $\mathbf{p}_i^{k+1} (v)$ such that $\mathbf{p}_i^{k+1} (u)\ge \mathbf{p}_i^{k+1} (v)$. Then, the expectation term in (\ref{statement2}) has,
\begin{equation}
  \begin{aligned}
  \mathbb{E}\big[\|\tilde{\mathcal{G}}_\gamma\big(\nabla \hat{f}(P^k), P^k\big)\|^2\big]&=1/\gamma^2\cdot\mathbb{E}\big[\|P^k - P^{k+1}\|^2\big] \\
  &= 1/\gamma^2\cdot\mathbb{E}\bigg[\sum_{i=1}^I\|\mathbf{p}_i^k - \mathbf{p}_i^{k+1}\|^2\bigg]\\
  &\ge 1/\gamma^2\cdot\mathbb{E}\big[\|\mathbf{p}_i^k - \mathbf{p}_i^{k+1}\|^2\big].
\end{aligned}
\end{equation}
According to the iteration of Algorithm~\ref{algo:SGD} and the computation~(\ref{plus}) of the projection $\Pi_\mathcal{S}$, then we have
\begin{align}
    \begin{cases}
     \mathbf{p}_i^{k+1} (u) = {\mathbf{p}}_{i}^k(u) + \gamma\cdot F_{-i}(u) - \lambda;\\
     \mathbf{p}_i^{k+1} (v) = {\mathbf{p}}_{i}^k(v) + \gamma\cdot F_{-i}(v) - \lambda.
  \end{cases}
\end{align}
Note that $F_{-i}(u)$ and $F_{-i}(v)$ are the $u$-th and $v$-th components of the sampled gradient $\nabla \hat{f}(P^k)$. {Since we have assumed that $\mathbf{p}_i^{k+1} (u)\ge \mathbf{p}_i^{k+1} (v) >0$, it can be verified that $F_{-i}(u)$ has to be the maximum one against all other $F_{-i}(n)$'s. Next, based on {the maximum distinguishable assumption}}, we know that $\mathbb{E}[F_{-i}(u)]$ has to be strictly greater than $ \mathbb{E}[F_{-i}(v)]$. Then, let $\Delta = \mathbb{E}[F_{-i}(u)] - \mathbb{E}[F_{-i}(v)] >0$, it holds that
\begin{equation}\label{dist}
  \begin{aligned}
   &\mathbb{E}\big[\|\mathbf{p}_i^k - \mathbf{p}_i^{k+1}\|^2 \big]\\
   &\hspace{-4pt}\overset{(2a)}{\ge} \|\mathbb{E}\big[\mathbf{p}_i^k - \mathbf{p}_i^{k+1}\big]\|^2\\
   &\ge \big(\gamma \cdot \mathbb{E}[F_{-i}(u)] - \mathbb{E}[\lambda]\big)^2 + \big(\gamma \cdot \mathbb{E}[F_{-i}(v)] - \mathbb{E}[\lambda]\big)^2\\
   & = \frac{1}{2}\big(2\gamma  \mathbb{E}[F_{-i}(v)]+ \gamma\Delta -2\mathbb{E}[\lambda]\big)^2 + \frac{1}{2} \gamma^2\Delta^2\\
   & \ge \frac{1}{2} \gamma^2\Delta^2.
 \end{aligned} 
\end{equation}
Note that $(2a)$ follows from the Jensen's inequality. Thus, the proof of statement 1) is completed.

\textit{Proof of statement 2)}:
According to the above Lemma~\ref{lemma:vertices}, we know that $P^{k+1}$ must be not the collection of simplex vertices, if the step-size $\gamma$ is choose under the condition and $P^k$ is the collection of vertices. In other words, $P^k \in \bar{\mathcal{S}}^k$ implies $P^{k+1} \in \bar{\mathcal{S}}_c^{k+1}$, and conversely, $P^{k+1} \in \bar{\mathcal{S}}^{k+1}$ implies $P^{k} \in \bar{\mathcal{S}}_c^k$. Therefore, the proof of statement 2) can be done by following exactly the same path of statement 1).

Now, recall that the stochastic gradient $\nabla \hat{f}(P^k)$ is \textit{i.i.d.} sampled with the sample-size $M$. Let us denote $\nabla \hat{f}_s(P^k)$ as the gradient decided by one single sample $s = 1,2,\cdots, M$, thus we know
\begin{equation}
  \begin{aligned}
    &\mathbb{E}\big[\|\nabla \hat{f}(P^k) - \nabla {f}(P^k)\|^2 \big] \\
    & = \mathbb{E}\bigg[\Big\|\frac{1}{M} \sum_{s=1}^M\big(\nabla \hat{f}_s(P^k) - \nabla {f}(P^k)\big)\Big\|^2 \bigg]\\
    & = \frac{1}{M^2}\cdot \sum_{s=1}^M\mathbb{E}\Big[\| \nabla \hat{f}_s(P^k) - \nabla {f}(P^k)\|^2 \Big].
  \end{aligned}
\end{equation}
Furthermore, by the definition of the function $f(P)$, it can be immediately verified that its gradient $\nabla {f}(P^k)$ is always bounded for $\forall P^k$, so is the \textit{i.i.d.} sampled stochastic gradient $\hat{f}_s(P^k)$. Based on this, we can have that the variance term $\mathbb{E}\big[\| \nabla \hat{f}_s(P^k) - \nabla {f}(P^k)\|^2 \big]$ is bounded. Therefore, the above two statements can further imply the following two conditions: there exists a constant $B_1$ such that, 
\begin{enumerate}
    \item if it is known that $P^{k+1} \in \bar{\mathcal{S}}_c^{k+1}$, then
  \begin{equation}\label{cond1}
      \begin{aligned}
   & \mathbb{E}\big[\|\nabla \hat{f}(P^k) - \nabla {f}(P^k)\|^2 \big] \\
   &\le B_1/M \cdot \mathbb{E}\big[\|\tilde{\mathcal{G}}_\gamma\big(\nabla \hat{f}(P^k), P^k\big)\|^2\big];
  \end{aligned}
  \end{equation}

    \item if it is known that $P^{k} \in \bar{\mathcal{S}}^{k}$, then
  \begin{equation}\label{cond2}
      \begin{aligned}
   &\hspace{-20pt}\mathbb{E}\big[\|\nabla \hat{f}(P^{k-1}) \hspace{-2pt}-\hspace{-2pt} \nabla {f}(P^{k-1})\|^2  +\|\nabla\hat{f}(P^k)\hspace{-2pt}-\hspace{-2pt} \nabla {f}(P^k)\|^2 \big] \\
   &\hspace{-20pt}\le B_1/M \cdot \mathbb{E}\big[\|\tilde{\mathcal{G}}_\gamma\big(\nabla \hat{f}(P^k), P^k\big)\|^2\big].
  \end{aligned}
  \end{equation}
\end{enumerate}

With the help of the above two inequalities (\ref{cond1}) and (\ref{cond2}), we are now ready to prove the statement in the lemma. Let us first denote $q_k$ the probability that $P^k$ is a collection of simplex vertices, i.e., 
 \begin{align}
   q_k:= \text{Pr}(P^k \in \bar{\mathcal{S}}^k).
 \end{align}
For the notational convenience, we denote
\begin{align}\label{threeTerms}
  \begin{cases}
    \textbf{I}^k = \mathbb{E}\Big[\mathbb{E}\big[\|\tilde{\mathcal{G}}_\gamma\big(\nabla \hat{f}(P^k), P^k\big)\|^2\big]\,\big\vert\,P^k \in \bar{\mathcal{S}}^k \Big];\\
    \textbf{II}^k = \mathbb{E}\Big[\mathbb{E}\big[\|\tilde{\mathcal{G}}_\gamma\big(\nabla \hat{f}(P^k), P^k\big)\|^2\big]\,\big\vert\,P^k\hspace{-2pt} \in\bar{\mathcal{S}}_c^k, P^{k+1}\hspace{-2pt} \in\hspace{-2pt} \bar{\mathcal{S}}_c^{k+1}\Big];\\
    \textbf{III}^k =  \mathbb{E}\Big[\mathbb{E}\big[\|\tilde{\mathcal{G}}_\gamma\big(\nabla \hat{f}(P^k), P^k\big)\|^2\big]\,\big\vert\,P^{k+1} \in \bar{\mathcal{S}}^{k+1} \Big].
  \end{cases}
\end{align}
It should be remarked that, in (\ref{threeTerms}), while the inner expectation is taken with respect to the stochastic gradient $\nabla \hat{f}(P^k)$, the outer expectation is taken with respect to the randomness of $P^k$ and $P^{k+1}$. Consequently, it holds that, for $\forall k \ge 1$,
  \begin{equation}\label{one}
   \begin{aligned}
     &\mathbb{E}\big[\|\tilde{\mathcal{G}}_\gamma\big(\nabla \hat{f}(P^k), P^k\big)\|^2\big] \\
     &= q_k \cdot \textbf{I}^k + (1-q_k -q_{k+1})\cdot\textbf{II}^k + q_{k+1}\cdot \textbf{III}^k\\
     &\overset{(3a)}{\ge} q_kM/B_1 \cdot\mathbb{E}\big[\|\nabla \hat{f}(P^{k-1}) - \nabla {f}(P^{k-1})\|^2\big] \\
     & \hspace{10pt} +(1- q_{k+1})M/B_1 \cdot\mathbb{E}\big[\|\nabla \hat{f}(P^k) - \nabla {f}(P^k)\|^2 \big].
   \end{aligned}
 \end{equation}
Note that $(3a)$ is due to the inequalities (\ref{cond1}), (\ref{cond2}) and the fact that $\textbf{III}^k \ge 0$. According to (\ref{one}), we have
 \begin{equation}\label{finalInequality}
   \begin{aligned}
     &\sum_{k=0}^T\mathbb{E}\big[\|\tilde{\mathcal{G}}_\gamma\big(\nabla \hat{f}(P^k), P^k\big)\|^2\big] \\
     &\overset{(4a)}{=} (1-q_1)\cdot\textbf{II}^0 + q_{1}\cdot \textbf{III}^0\\
     & \hspace{10pt}+\sum_{k=1}^T \Big(q_k \cdot \textbf{I}^k + (1-q_k -q_{k+1})\cdot\textbf{II}^k + q_{k+1}\cdot \textbf{III}^k\Big)\\
     &\overset{(4b)}{\ge} (1-q_1)M/B_1\cdot\mathbb{E}\big[\|\nabla \hat{f}(P^{0}) - \nabla {f}(P^{0})\|^2\big] \\
     &\hspace{10pt}+\sum_{k=1}^T q_kM/B_1 \cdot\mathbb{E}\big[\|\nabla \hat{f}(P^{k-1}) - \nabla {f}(P^{k-1})\|^2\big] \\
     &\hspace{10pt}+\sum_{k=1}^T (1- q_{k+1})M/B_1 \cdot\mathbb{E}\big[\|\nabla \hat{f}(P^k) - \nabla {f}(P^k)\|^2 \big]\\
     & = M/B_1 \cdot \sum_{k = 0}^T \mathbb{E}\big[\|\nabla \hat{f}(P^k) - \nabla {f}(P^k)\|^2 \big] \\
     & \hspace{10pt} - q_{T+1}M/B_1 \cdot\mathbb{E}\big[\|\nabla \hat{f}(P^T) - \nabla {f}(P^T)\|^2 \big]\\ 
     & \overset{(4c)}{\ge} M/B_1 \cdot \sum_{k = 0}^T \mathbb{E}\big[\|\nabla \hat{f}(P^k) - \nabla {f}(P^k)\|^2 \big] - MB_0/B_1.
   \end{aligned}
 \end{equation}
 Note that $(4a)$ is due to the fact that the initialization $P^0$ is not the collection of vertices, i.e. $q_0 = 0$; $(4b)$ comes from the inequality (\ref{one}); and $(4c)$ is based on the fact that the variance term $\mathbb{E}\big[\|\nabla \hat{f}(P^T) - \nabla {f}(P^T)\|^2 \big]$ can be upper bounded by the constant $B_0$. Rearranging the inequality (\ref{finalInequality}) and noticing the definition (\ref{compactAlgo}) of the stacked gradient mapping complete the proof of the lemma.
\end{IEEEproof}

After showing the above lemmas, we are now in the position to prove the theorem. Since the Hessian of the function $f(P)$ is always bounded, it can be immediately verified that the gradient of $f(P)$ is Lipschitz continuous, so is the gradient of $-f(P)$. Thus, there exists a constant $L >0$ such that,
\begin{equation}\label{mainProof}
  \begin{aligned}
  &-f(P^{k+1}) \\
  &\le -f(P^k) + \big\langle \hspace{-2pt}-\hspace{-2pt}\nabla f(P^k),\;P^{k+1} - P^k\big\rangle +\frac{L}{2}\|P^{k+1} - P^k\|^2\\
  &\overset{(5a)}{=} -f(P^k) +\gamma\Big\langle\nabla f(P^k) \pm \nabla \hat{f}(P^k),\;\tilde{\mathcal{G}}_\gamma\big(\nabla \hat{f}(P^k),P^k\big)\Big\rangle\\
    & \quad\;+ \frac{\gamma^2L}{2}\|\tilde{\mathcal{G}}_\gamma\big(\nabla \hat{f}(P^k),P^k\big)\|^2\\
  &\overset{(5b)}{\le} -f(P^k) + (\frac{\gamma^2L}{2} - \gamma)\|\tilde{\mathcal{G}}_\gamma\big(\nabla \hat{f}(P^k),P^k\big)\|^2\\
    & \quad\;+\gamma\Big\langle\nabla f(P^k) -\nabla \hat{f}(P^k),\\
    & \quad\quad\quad\quad\tilde{\mathcal{G}}_\gamma\big(\nabla \hat{f}(P^k),P^k\big) \pm\tilde{\mathcal{G}}_\gamma\big(\nabla {f}(P^k),P^k\big) \Big\rangle\\
  &\overset{(5c)}{\le} -f(P^k) + (\frac{\gamma^2L}{2}- \gamma)\|\tilde{\mathcal{G}}_\gamma\big(\nabla \hat{f}(P^k),P^k\big)\|^2\\
  & \quad\;+ \gamma\Big\langle\nabla f(P^k) - \nabla \hat{f}(P^k),\;\tilde{\mathcal{G}}_\gamma\big(\nabla {f}(P^k),P^k\big) \Big\rangle\\
    & \quad\;+ \gamma\|\nabla f(P^k) -\nabla \hat{f}(P^k)\|^2.
  \end{aligned}
\end{equation}
Note that ($5a$) is due to the definition of gradient mapping; ($5b$)~comes from Lemma~\ref{lemma:GradMapping1}; and ($5c$) is due to the Cauchy-Schwartz inequality and Lemma~\ref{lemma:GradMapping2}. 
Now, let us take expectation on the inequality (\ref{mainProof}), with respect to the random sampling of stochastic gradient $\nabla \hat{f}(P^k)$ by given the probability distribution $P^k$. Since $\nabla \hat{f}(P^k)$ is the unbiased estimation of the full gradient $\nabla f(P^k)$ according to Lemma~\ref{lemma:unbiasedGrad}, it holds that
\begin{equation}\label{eq4}
  \begin{aligned}
     \mathbb{E}\big[f(P^{k+1})\big] \hspace{-2pt}- f(P^k)\ge& (\gamma \hspace{-2pt}-\hspace{-2pt}\frac{\gamma^2L}{2})\cdot\mathbb{E}\big[\|\tilde{\mathcal{G}}_\gamma\big(\nabla \hat{f}(P^k),P^k\big)\|^2\big]\\
    & - \gamma\cdot\mathbb{E} \big[\|\nabla f(P^k) -\nabla \hat{f}(P^k)\|^2\big].
  \end{aligned}
\end{equation}
Consequently, summing up the above inequality (\ref{eq4}) for all $0 \le k \le T$ and taking the expectation with respect to the random sampling for all iterations, we have
\begin{equation}\label{eq3}
  \begin{aligned}
  &\mathbb{E}\big[f(P^{T+1})\big] - f(P^0)\\
  &\ge(\gamma -\frac{\gamma^2L}{2})\cdot\sum_{k = 0}^T \mathbb{E}\big[\|\tilde{\mathcal{G}}_\gamma\big(\nabla \hat{f}(P^k),P^k\big)\|^2\big] \\
  &\quad\;- \gamma\cdot \sum_{k = 0}^T\mathbb{E} \big[\|\nabla f(P^k) -\nabla \hat{f}(P^k)\|^2\big]\\
  &\overset{(6a)}{\ge} -\gamma B_0 + (\gamma \hspace{-2pt}-\hspace{-2pt} \frac{B_1\gamma}{M}\hspace{-2pt}-\hspace{-2pt}\frac{\gamma^2L}{2} )\cdot\sum_{k = 0}^T \mathbb{E}\big[\|\tilde{\mathcal{G}}_\gamma\big(\nabla \hat{f}(P^k),P^k\big)\|^2\big].
\end{aligned}
\end{equation}
Note that $(6a)$ follows from Lemma~\ref{lemma:boundedVar}. Now, suppose that $P^\star$ is the optimal solution for solving problem~(\ref{prob_max}), i.e., \mbox{$\mathbb{E}\big[f(P^{T+1})\big] \le f(P^\star),\, \forall \, T\in \mathbb{R}_+$}. Then, the above inequality~(\ref{eq3}) implies that, if the sample-size $M$ and step-size $\gamma$ are chosen satisfying  $M > B_1$ and $\gamma -B_1\gamma/M- {\gamma^2L}/{2} > 0$, the non-negative sequence $\big\{\mathbb{E}\big[\|\tilde{\mathcal{G}}_\gamma\big(\nabla \hat{f}(P^k),P^k\big)\|^2\big]\big\}_{k \in \mathbb{N}_+}$ is~summable, i.e.,
\begin{align}
  \sum_{k = 0}^\infty \mathbb{E}\big[\|\tilde{\mathcal{G}}_\gamma\big(\nabla \hat{f}(P^k),P^k\big)\|^2\big] \le \frac{f(P^\star) - f(P^0) + \gamma B_0}{\gamma -B_1\gamma/M- {\gamma^2L}/{2}}.
\end{align}
Thus, $\mathbb{E}\big[\|\tilde{\mathcal{G}}_\gamma\big(\nabla \hat{f}(P^k),P^k\big)\|^2\big]$ converges to zero; and furthermore, its running average converges at the rate of $\mathcal{O}(1/T)$.

\subsection{Proof of Proposition~\ref{prop:deGradMapping}}

Let us first remark that the condition~(\ref{deGradMapping}) simply implies that the following equation holds for all $k \le t \le k+D-1$,
  \begin{align}\label{deGradMapping_single}
  \mathbb{E} \Big[\big\|\mathbf{p}_i^{t} - \Pi_{\mathcal{S}}\big( \mathbf{p}_i^{t} + \gamma \cdot \nabla_{\mathbf{p}_i} \hat{f}_\delta(P_i^{t-}) \big)\big\|^2\Big] = 0,\; \forall i \in \mathcal{I}.
  \end{align}
According to the iteration (\ref{delayedSGD}), it can be immediately verified that $P^{t+1} = P^t$ is true for all $k \le t \le k+D-2$. Therefore, to prove the statement in Proposition~\ref{prop:deGradMapping}, it will suffice to show that $P^{k+D-1}$ is an equilibrium solution.

Since each $P^t$ is identical within the entire time-window $k \le t \le k+D-1$, then let $t = k+D-1$, the equation~(\ref{deGradMapping_single}) implies that, for all $i\in \mathcal{I}$,
  \begin{align}\label{deGradMapping_nodelay}
  \mathbb{E} \Big[\big\|\mathbf{p}_i^{k+D -1} - \Pi_{\mathcal{S}}\big( \mathbf{p}_i^{k+D -1} + \gamma \cdot \nabla_{\mathbf{p}_i} \hat{f}(P^{k+D -1}) \big)\big\|^2\Big] = 0.
  \end{align}
Note that, in (\ref{deGradMapping_nodelay}), the stochastic gradient $\nabla_{\mathbf{p}_i} \hat{f}(P^{k+D -1})$ is evaluated without time-delays. As a result of the above Proposition~\ref{prop:gradMapping}, we know that $P^{k+D -1}$ has to be an equilibrium. Therefore, the proof is completed.
\subsection{Proof of Theorem~\ref{theorem:delay}}

As similar to the previous proof, let us first mention that the iteration (\ref{delayedSGD}) of Algorithm~2 can be compactly expressed~as,
\begin{equation}\label{compactDelayAlgo}
  \begin{aligned}
        P^{k+1}= P^k - \gamma\cdot\tilde{\mathcal{G}}_\gamma\big(\nabla \hat{f}_\delta(\PsD^{k-}),P^k\big),
  \end{aligned}
\end{equation}
where $\tilde{\mathcal{G}}_\gamma\big(\nabla \hat{f}_\delta(\PsD^{k-}),P^k\big)$ is the stacked gradient mapping and $\PsD^{k-} = [P_i^{k-}]_{i \in \mathcal{I}}$ collects the delayed distributions $P_i^{k-}$ for all $i \in \mathcal{I}$. Now, according to the Lipschitz continuous gradient of the function$-f(P)$, we invoke the descent lemma again and it holds that,
\begin{equation}\label{descent}
  \begin{aligned}
  &-f(P^{k+1}) \\
  &\le -f(P^k) + \big\langle \hspace{-2pt}-\hspace{-2pt}\nabla f(P^k),\;P^{k+1} - P^k\big\rangle +\frac{L}{2}\|P^{k+1} - P^k\|^2\\
  &\overset{}{=} -f(P^k) +\gamma\Big\langle \nabla \hat{f}_\delta(\PsD^{k-}),\;\tilde{\mathcal{G}}_\gamma\big(\nabla \hat{f}_\delta(\PsD^{k-}),P^k\big)\Big\rangle\\
    & \quad\; + \gamma\Big\langle\nabla f(P^k) - \nabla \hat{f}_\delta(\PsD^{k-}),\;\tilde{\mathcal{G}}_\gamma\big(\nabla \hat{f}_\delta(\PsD^{k-}),P^k\big)\Big\rangle\\
    & \quad\; + \frac{\gamma^2L}{2}\|\tilde{\mathcal{G}}_\gamma\big(\nabla \hat{f}_\delta(\PsD^{k-}),P^k\big)\|^2\\
  &\overset{(7a)}{\le} -f(P^k) + (\frac{\gamma^2L}{2} - \gamma)\|\tilde{\mathcal{G}}_\gamma\big(\nabla \hat{f}_\delta(\PsD^{k-}),P^k\big)\|^2\\
    & \quad\; + \gamma\Big\langle\nabla f(P^k) - \nabla {f}_\delta(\PsD^{k-}),\;\tilde{\mathcal{G}}_\gamma\big(\nabla \hat{f}_\delta(\PsD^{k-}),P^k\big)\Big\rangle\\
    & \quad\; + \gamma\Big\langle\nabla {f}_\delta(\PsD^{k-}) - \nabla \hat{f}_\delta(\PsD^{k-}),\;\tilde{\mathcal{G}}_\gamma\big(\nabla \hat{f}_\delta(\PsD^{k-}),P^k\big)\Big\rangle.
  \end{aligned}
\end{equation}
Note that the inequality $(7a)$ is due to Lemma~\ref{lemma:GradMapping1}; furthermore, we use $\nabla {f}_\delta(\PsD^{k-})$ to denote the full gradient which is based on the delayed probability distributions $\PsD^{k-}$. It should be emphasized that the sampled gradient $\nabla \hat{f}_\delta(\PsD^{k-})$ is an unbiased estimation of the full gradient $\nabla {f}_\delta(\PsD^{k-})$. Thus, according to the Cauchy-Schwartz inequality and Lemma~\ref{lemma:GradMapping2}, the above (\ref{descent}) can be continued as 
\begin{equation}\label{descent2}
  \begin{aligned}
  &-f(P^{k+1}) \\[-5pt]
  &{\le} -f(P^k) + (\frac{\gamma^2L}{2} - \gamma)\|\tilde{\mathcal{G}}_\gamma\big(\nabla \hat{f}_\delta(\PsD^{k-}),P^k\big)\|^2\\
    & \quad\; + \gamma\Big\langle\nabla f(P^k) - \nabla {f}_\delta(\PsD^{k-}),\;\tilde{\mathcal{G}}_\gamma\big(\nabla \hat{f}_\delta(\PsD^{k-}),P^k\big)\Big\rangle\\
    & \quad\;+\gamma\Big\langle\nabla {f}_\delta(\PsD^{k-}) - \nabla \hat{f}_\delta(\PsD^{k-}),\\
    & \quad\quad\quad\quad\tilde{\mathcal{G}}_\gamma\big(\nabla \hat{f}_\delta(\PsD^{k-}),P^k\big) \pm\tilde{\mathcal{G}}_\gamma\big(\nabla {f}_\delta(\PsD^{k-}),P^k\big) \Big\rangle\\
  &{\le} -f(P^k) + (\frac{\gamma^2L}{2} - \gamma)\|\tilde{\mathcal{G}}_\gamma\big(\nabla \hat{f}_\delta(\PsD^{k-}),P^k\big)\|^2\\
    & \quad\; + \gamma\big\|\nabla f(P^k) - \nabla {f}_\delta(\PsD^{k-})\big\|\big\|\tilde{\mathcal{G}}_\gamma\big(\nabla \hat{f}_\delta(\PsD^{k-}),P^k\big)\big\|\\
    & \quad\; + \gamma\Big\langle\nabla {f}_\delta(\PsD^{k-}) - \nabla \hat{f}_\delta(\PsD^{k-}),\;\tilde{\mathcal{G}}_\gamma\big(\nabla {f}_\delta(\PsD^{k-}),P^k\big) \Big\rangle\\
    & \quad\; + \gamma\big\|\nabla {f}_\delta(\PsD^{k-}) - \nabla \hat{f}_\delta(\PsD^{k-})\big\|^2.
  \end{aligned}
\end{equation}

Now, taking the expectation on both sides and summing up the inequalities for all {$0 \le k \le T$}, it holds that
\begin{equation}\label{mainDelay}
  \begin{aligned}
  &\mathbb{E}\big[f(P^{T+1})\big] - f(P^0)\\
  &\ge(\gamma -\frac{\gamma^2L}{2})\cdot\sum_{k = 0}^T \mathbb{E}\big[\|\tilde{\mathcal{G}}_\gamma\big(\nabla \hat{f}_\delta(\PsD^{k-}),P^k\big)\|^2\big] \\
  &\quad\;- \gamma\cdot \underbrace{\sum_{k = 0}^T\mathbb{E} \big[\|\nabla {f}_\delta(\PsD^{k-}) - \nabla \hat{f}_\delta(\PsD^{k-})\|^2\big]}_{\hspace{15pt}: =\mathcal{T}_1}\\
  &- \gamma\cdot \underbrace{\sum_{k = 0}^T\mathbb{E} \big[\|\nabla f(P^k) - \nabla {f}_\delta(\PsD^{k-})\|\|\tilde{\mathcal{G}}_\gamma\big(\nabla \hat{f}_\delta(\PsD^{k-}),P^k\big)\|}_{\hspace{15pt}: =\mathcal{T}_2}\big].
\end{aligned}
\end{equation}
As shown in the above inequality, let us denote the last two summation terms as $\mathcal{T}_1$ and~$\mathcal{T}_2$, respectively. Next, we prove the following two lemmas which upper bound $\mathcal{T}_1$ and~$\mathcal{T}_2$ by the summation of gradient mappings.

\begin{lemma}\label{Lemma7}
  Suppose that the sequence $\{P^k\}_{k \in \mathbb{N}_+}$ is the set of iterates generated by Algorithm~2 and the initialization $P^0$ is not a collection of simplex vertices. {Let the step-size~$\gamma$ satisfy the condition $\gamma < 2/\Delta^\text{max}$}, then there exist constants $C_0>0$ and $ C_1> 0$ such that the following holds,
\begin{equation}
  \begin{aligned}
\mathcal{T}_1 \le C_0 + C_1/M \cdot\sum_{k=0}^T\mathbb{E}\Big[\|\tilde{\mathcal{G}}_\gamma\big(\nabla \hat{f}_\delta(\PsD^{k-}), P^k\big)\|^2\Big].
\end{aligned}
\end{equation}
\end{lemma}
\begin{IEEEproof}
  This proof can be done by following exactly the similar path of Lemma~\ref{lemma:boundedVar}, and thus we omit the details.
\end{IEEEproof}

\begin{lemma}\label{Lemma8}
  Suppose that the conditions on Lemma~\ref{Lemma7} are satisfied, then there exist constants $C_2>0$ and $C_3 >0$ such that the following holds,
\begin{equation}
  \begin{aligned}
\mathcal{T}_2 \le \gamma C_2 + \gamma C_3 \cdot\sum_{k=0}^T\mathbb{E}\Big[\|\tilde{\mathcal{G}}_\gamma\big(\nabla \hat{f}_\delta(\PsD^{k-}), P^k\big)\|^2\Big].
\end{aligned}
\end{equation}
\end{lemma}
\begin{IEEEproof}
  We first recall that $\nabla {f}_\delta(\PsD^{k-})$ represents the stacked full gradient with respect to the delayed probability distributions $\PsD^{k-}$. Precisely, let us denote
  \begin{align}
    \nabla {f}_\delta(\PsD^{k-}) = [\nabla_{\mathbf{p}_i} {f}(P_i^{k-})]_{i \in \mathcal{I}},
  \end{align}
  where $P_i^{k-}$ captures all the delayed distributions associated with the $i$-th agent. On this account, we can have
  \begin{equation}\label{biasTerm}
    \begin{aligned}
      &\|\nabla f(P^k) - \nabla {f}_\delta(\PsD^{k-})\|^2 \\
      &= \sum_{i=1}^I \|\nabla_{\mathbf{p}_i} {f}(P^k) -\nabla_{\mathbf{p}_i} {f}(P_i^{k-})\|^2 \\
      & \overset{(8a)}{\le} \sum_{i=1}^I L_i^2 \|P^k -P_{i}^{k-}\|^2 \\
      & = \sum_{i=1}^I\sum_{j\neq i}^I  L_i^2  \|\mathbf{p}_j^k -\mathbf{p}_j^{k-\tau_{ij}}\|^2\\
      &\overset{(8b)}{\le} \sum_{i=1}^I\sum_{j\neq i}^I  L_i^2 \sum_{t = 0}^{\tau_{ij}-1} \|\mathbf{p}_j^{k-t} -\mathbf{p}_j^{k-t-1}\|^2,
    \end{aligned}
  \end{equation}
  where $(8a)$ is due to the fact that each gradient $\nabla_{\mathbf{p}_i} {f}(P)$ is $L_i$-Lipschitz continuous and $(8b)$ comes from the triangle inequality. In addition, according to Lemma~\ref{lemma:GradMapping1} and Cauchy-Schwartz inequality, it holds that
    \begin{equation}
    \begin{aligned}
& \|{\mathcal{G}}_\gamma\big(\nabla_{\mathbf{p}_i} \hat{f}_\delta(P_i^{k-}), \mathbf{p}_i^k\big)\|^2 \\
&\le - \Big\langle \nabla_{\mathbf{p}_i} \hat{f}_\delta(P_i^{k-}),\;{\mathcal{G}}_\gamma\big(\nabla_{\mathbf{p}_i} \hat{f}_\delta(P_i^{k-}), \mathbf{p}_i^k\big)\Big\rangle\\
      & \le \|\nabla_{\mathbf{p}_i} \hat{f}_\delta(P_i^{k-})\|\cdot\|{\mathcal{G}}_\gamma\big(\nabla_{\mathbf{p}_i} \hat{f}_\delta(P_i^{k-}), \mathbf{p}_i^k\big)\|,
    \end{aligned}
  \end{equation}
and thus for $\forall k \in \mathbb{N}_+$,
    \begin{equation}\label{diff}
    \begin{aligned}
\|\mathbf{p}_i^{k+1} -\mathbf{p}_i^{k} \|^2 &= \gamma^2\|{\mathcal{G}}_\gamma\big(\nabla_{\mathbf{p}_i} \hat{f}_\delta(P_i^{k-}), \mathbf{p}_i^k\big)\|^2 \\
      & \le \gamma^2 \|\nabla_{\mathbf{p}_i} \hat{f}_\delta(P_i^{k-})\|^2.
    \end{aligned}
  \end{equation}
Note that the above inequality also shows that the gradient mapping is always bounded by the stochastic gradient.
Next, based on the inequalities (\ref{biasTerm}), (\ref{diff}) and the facts that the gradient $\nabla_{\mathbf{p}_i} \hat{f}_\delta(P_i^{k-})$ is bounded and $\tau_{ij} \le D,\; \forall i,j \in \mathcal{I}$, we know that there must exist a constant $\beta > 0$ such that
  \begin{equation}\label{biasTerm2}
    \begin{aligned}
      &\|\nabla f(P^k) - \nabla {f}_\delta(\PsD^{k-})\| \le \gamma \beta.
    \end{aligned}
  \end{equation}
Consequently, the summation term $\mathcal{T}_2$ can be bounded by
\begin{align}\label{final1}
  \mathcal{T}_2 \le \gamma \beta \cdot \sum_{k = 0}^T\mathbb{E} \big[\|\tilde{\mathcal{G}}_\gamma\big(\nabla \hat{f}_\delta(\PsD^{k-}),P^k\big)\| \big].
\end{align}

Now, the rest of the proof follows the similar path of the proof in Lemma \ref{lemma:boundedVar} (or Lemma~\ref{Lemma7}). 
It can be shown that there exist two constants $\rho_0 > 0$ and $\rho_1 > 0$ such that
\begin{equation}\label{final2}
  \begin{aligned}
         &\sum_{k=0}^T\mathbb{E}\big[\|\tilde{\mathcal{G}}_\gamma\big(\nabla \hat{f}_\delta(\PsD^{k-}), P^k\big)\|^2\big] \\
         &\ge \frac{1}{\rho_1}\cdot\sum_{k=0}^T\mathbb{E}\big[\|\tilde{\mathcal{G}}_\gamma\big(\nabla \hat{f}_\delta(\PsD^{k-}), P^k\big)\|\big] -\frac{\rho_0}{\rho_1}.
  \end{aligned}
\end{equation}
Combining the inequalities (\ref{final1}) and (\ref{final2}), we can have
\begin{align}
  \mathcal{T}_2 \le \gamma \beta \rho_0 + \gamma \beta \rho_1 \cdot\sum_{k=0}^T\mathbb{E}\Big[\|\tilde{\mathcal{G}}_\gamma\big(\nabla \hat{f}_\delta(\PsD^{k-}), P^k\big)\|^2\Big].
\end{align}
Therefore, let $C_2 = \beta \rho_0$ and $C_3 = \beta \rho_1$ respectively, the proof is completed.
\end{IEEEproof}

Next, we prove the statement in Theorem~\ref{theorem:delay}. Taking into account Lemma~\ref{Lemma7} and Lemma~\ref{Lemma8} together, the inequality~(\ref{mainDelay}) can be continued as 
\begin{equation}\label{final}
  \begin{aligned}
  &\mathbb{E}\big[f(P^{T+1})\big] - f(P^0)\\
  &\ge(\gamma -\frac{\gamma C_1}{M} - \gamma^2 C_3-\frac{\gamma^2L}{2})\cdot\sum_{k = 0}^T \mathbb{E}\big[\|\tilde{\mathcal{G}}_\gamma\big(\nabla \hat{f}_\delta(\PsD^{k-}),P^k\big)\|^2\big] \\
  & \quad\;- \gamma C_0 - \gamma^2 C_2.
\end{aligned}
\end{equation}
As a result, it holds that,
\begin{equation}
    \begin{aligned}
  &\sum_{k = 0}^\infty \mathbb{E}\big[\|\tilde{\mathcal{G}}_\gamma\big(\nabla \hat{f}_\delta(\PsD^{k-}),P^k\big)\|^2\big] \\
  &\le \frac{f(P^\star) - f(P^0) + \gamma C_0 + \gamma^2 C_2}{\gamma -{\gamma C_1}/{M} - \gamma^2 C_3-{\gamma^2L}/{2}}.
\end{aligned}
\end{equation}
Therefore, if the sample-size $M$ and step-size $\gamma$ are chosen satisfying  $M > C_1$ and $\gamma -{\gamma C_1}/{M} - \gamma^2 C_3-{\gamma^2L}/{2} > 0$, then we can have that $\mathbb{E}\big[\|\tilde{\mathcal{G}}_\gamma\big(\nabla \hat{f}(P^k),P^k\big)\|^2\big]$ converges to zero; and furthermore, its running average converges at the rate of $\mathcal{O}(1/T)$.

\bibliographystyle{unsrt}
\bibliography{main}

\end{document}